\def\jcap{JCAP}
\def\mnras{MNRAS}             
\def\aap{A\&A}             
\def\apjs{ApJS}             
\def\apjl{ApJL}             
\def\baas{BAAS}
\def\araa{ARA\&A}
\def\aj{AJ}
\def\pasj{PASJ}

\documentclass[%
reprint,
superscriptaddress,
nofootinbib,
nolongbibliography,
 amsmath,amssymb,
 aps,
prd,
floatfix,
]{revtex4-2}

\usepackage{graphicx}
\usepackage{dcolumn}
\usepackage{bm}
\usepackage{xcolor}

\definecolor{maroon}{RGB}{192,0,0}
\usepackage[colorlinks=true, citecolor=maroon,
linkcolor=maroon,
urlcolor=blue]{hyperref}
\usepackage{times}

\usepackage{amsmath}
\usepackage{array,multirow}
\usepackage{cancel}
\newcommand{\diffop}{\mathrm{d}}

\newcommand{\mmat}[1]{{\mathbf{#1}}}
\newcommand{\mP}{\mathcal{P}}

\newcommand{\mL}{\mathcal{L}}
\newcommand{\ia}{\text{IA}}
\newcommand{\mvec}[1]{{\bm{#1}}}
\newcommand{\md}{\mvec{d}}
\newcommand{\mtheta}{\mvec{\theta}}

\newcommand{\cosmolike}{{\sc cosmolike}~}

\newcommand{\red}[1]{{\color{red} #1}}

\begin{document}

\title{Machine Learning LSST 3$\times$2pt analyses - forecasting the impact of systematics on cosmological constraints using neural networks}

\author{Supranta S. Boruah}
 \email{supranta@sas.upenn.edu}
\affiliation{Department of Astronomy and Steward Observatory, University of Arizona, 933 N Cherry Ave, Tucson, AZ 85719, USA 
}%
\affiliation{Department of Physics and Astronomy, University of Pennsylvania, Philadelphia, PA 19104, USA}
\author{Tim Eifler}
\affiliation{Department of Astronomy and Steward Observatory, University of Arizona, 933 N Cherry Ave, Tucson, AZ 85719, USA 
}%
\author{Vivian Miranda}
\affiliation{Department of Physics \& Astronomy, Stony Brook University, Stony Brook, NY 11794, USA}
\affiliation{C. N. Yang Institute for Theoretical Physics, Stony Brook University, Stony Brook, NY, 11794, USA}
\author{Elyas Farah}
\affiliation{Department of Astronomy and Steward Observatory, University of Arizona, 933 N Cherry Ave, Tucson, AZ 85719, USA 
}%
\author{Jay Motka}
\affiliation{Department of Astronomy and Steward Observatory, University of Arizona, 933 N Cherry Ave, Tucson, AZ 85719, USA 
}%
\author{Elisabeth Krause}
\affiliation{Department of Astronomy and Steward Observatory, University of Arizona, 933 N Cherry Ave, Tucson, AZ 85719, USA 
}%
\affiliation{Department of Physics, University of Arizona, 1118 E. Fourth Street, Tucson, AZ, 85721, USA}
\author{Xiao Fang}
\affiliation{Berkeley Center for Cosmological Physics, UC Berkeley, CA 94720, USA 
}%
\affiliation{Department of Astronomy and Steward Observatory, University of Arizona, 933 N Cherry Ave, Tucson, AZ 85719, USA }
\author{Paul Rogozenski}
\affiliation{Department of Physics, University of Arizona, 1118 E. Fourth Street, Tucson, AZ, 85721, USA}
\author{The LSST Dark Energy Science Collaboration}

\date{\today}

\begin{abstract}
Validating modeling choices through simulated analyses and quantifying the impact of different systematic effects will form a major computational bottleneck in the preparation for 3$\times$2 analysis with Stage-IV surveys such as Vera Rubin Observatory's Legacy Survey of Space and Time (LSST). We can  significantly reduce the computational requirements by using machine learning based emulators, which allow us to run fast inference while maintaining the full realism of the data analysis pipeline. In this paper, we use such an emulator to run simulated 3$\times$2 (cosmic shear, galaxy-galaxy lensing, and galaxy clustering) analyses for mock LSST-Y1/Y3/Y6/Y10 surveys and study the impact of various systematic effects (galaxy bias, intrinsic alignment, baryonic physics, shear calibration and photo-$z$ uncertainties). Closely following the DESC Science Requirement Document (with several updates) our main findings are: {\it a)} The largest contribution to the `systematic error budget' of LSST 3$\times$2 analysis comes from galaxy bias uncertainties, while the contribution of baryonic and shear calibration uncertainties are significantly less important. {\it b)} Tighter constraints on intrinsic alignment and photo-$z$ parameters can improve cosmological constraints noticeably, which illustrates synergies of LSST and spectroscopic surveys. {\it c)} The scale cuts adopted in the DESC SRD may be too conservative and pushing to smaller scales can increase cosmological information significantly. {\it d)} We investigate the impact of photo-$z$ outliers on 3$\times$2 pt analysis and find that we need to determine the outlier fraction to within $5-10\%$ accuracy to ensure robust cosmological analysis. We caution that these findings depend on analysis choices (parameterizations, priors, scale cuts) and can change for different settings. 
\end{abstract}

\maketitle


\section{Introduction}\label{sec:intro}

The upcoming Vera Rubin Observatory's Legacy Survey of Space and Time \citep[LSST\footnote{\url{https://www.lsst.org/}},][]{LSST2019} will be part of a new era of cosmological surveys aimed at uncovering the nature of dark energy. Along with other Stage-IV surveys \citep{DETFreport2006} like Euclid \footnote{\url{https://sci.esa.int/web/euclid}} \citep{laa11}, the Nancy G. Roman Space Telescope \citep[NGRST\footnote{\url{https://roman.gsfc.nasa.gov/}},][]{sgb15}, and the Dark Energy Spectroscopic Instrument \citep[DESI \footnote{\url{https://www.desi.lbl.gov/}}][]{DESI16}, LSST will examine the nature of dark energy with an exquisite precision. Among the various probes of dark energy in LSST, multiprobe cosmological analyses with weak lensing, galaxy-galaxy lensing, and galaxy clustering, commonly referred to as `3$\times$2  analysis', will provide one of the strongest constraints on the dark energy parameters \citep{DESC_SRD}.

However, weak lensing and galaxy clustering measurements are strongly impacted by various systematic effects \citep{Mandelbaum2018}. The impact of systematic effects is mitigated through various strategies, for example, by cutting out relevant scales where the modeling is imprecise or by including additional `nuisance' parameters in the model. While these mitigation schemes can reduce the bias on cosmological parameters, using an aggressive scale cut or an overly flexible systematics model leads to a loss in constraining power. As a result, bias-variance tradeoff is at play when deciding our analysis choices that are applied to the actual dataset and mapping out different aspects of this tradeoff is crucial as we prepare for LSST data analysis. Such an exercise will allow us to prioritize algorithmic development or future observational efforts to mitigate the systematic effects most important for our cosmological analysis. Identifying priorities for systematics mitigation strategy is the objective of this paper.

The analysis choices for the Stage-III survey cosmic shear and multiprobe analyses are determined using the full data analysis pipeline to ensure that the results are robust to known systematic effects \citep[e.g.][]{Hikage2019, Heymans2021, Krause2021}. These impact studies require hundreds to thousands of computationally expensive Markov Chain Monte Carlo (MCMC) chains. Furthermore, the process needs to be repeated for each new probe or analysis. For Stage-IV survey data analysis, this challenge will become computationally even more expensive mostly because of the more complex systematics models that consequently span a larger parameter space. While simulated likelihood analysis with the full analysis pipeline is the most correct way to perform these impact studies, it is a serious computational bottleneck for future Stage-IV surveys. As such, we need robust forecasting pipelines that can perform fast, yet accurate inference.     

Fisher matrix formalism is a common technique for fast forecasting, but it assumes that the posterior can be described as a multivariate Gaussian \citep{Wolz2012} and it can show numerical instabilities in high dimensions \citep{Euclid2020, Bhandari2021, YahiaCherif2021}. Alternate proposals for fast cosmological analyses include analytically marginalizing subsets of the full parameter space \citep{Hadzhiyska2020,Stolzner2021, Hadzhiyska2023, RuizZapatero2023} but these methods too rely on the assumption that the impact of marginalized parameters on the data vectors can be linearized, which may not be suitable for nonlinear systematics models. Importance sampling is another approach frequently used for quick impact studies, but it only works close to the already sampled parameter space and it becomes problematic in high dimensional parameter spaces with degeneracies.

Machine-learning based emulators that are trained on the specific analysis pipeline are emerging as a computationally efficient and accurate avenue to predict the data vectors. However, standard emulator designs struggle to perform well in high-dimensions due to the curse of dimensionality. To circumvent this problem, iterative emulators have been proposed in the literature \citep{To2023, Boruah2022, Nygaard2022, Paranjape2022} where training points are acquired from a previous iteration of sampling. In this paper, we use the emulator introduced in \cite{Boruah2022} to extensively study different aspects of 3$\times$2  analysis with LSST. This method can perform accurate emulation in the high-dimensional parameter spaces of Stage-IV surveys such as LSST at a fraction of the computational cost of running the full pipeline. While we focus on LSST in this paper, our emulator provides a framework for undertaking the impact studies in preparation for any future photometric surveys such as Roman Space Telescope and Euclid. 

In this paper we study the impact of marginalization over different systematic effects closely following the Dark Energy Science Collaboration Science Requirement Document \citep[][DESC-SRD from hereon]{DESC_SRD}. The main differences/updates compared to the DESC-SRD are: 1) Our simulated likelihood analyses uses emulator-based MCMC runs, while the DESC-SRD uses Fisher matrix, 2) our data vector is comprised of real space angular correlation functions, while DESC-SRD performed its analysis in Fourier space, 3) we also marginalize over baryonic physics uncertainties, and 4) we chose a different lens sample and tomographic binning.

With respect to the latter point, the DESC-SRD uses the LSST Gold sample as the lens sample, whereas we perform a `lens=source' analysis where the source sample itself is used as the lens sample. Such analyses have been previously considered in \citep{Schaan2020, Fang2022} and are motivated by the idea that using one galaxy sample as both the lens and source sample reduces the number of systematic parameters in a likelihood analysis. For example, \cite{Fang2022} show an increase in the signal-to-noise of the data vectors and stronger self-calibration of the redshift distribution when going from the `standard' analysis to the `lens=source' analysis.

This paper is structured as follows: After describing the theoretical basics and LSST survey assumptions in section \ref{sec:background} and \ref{sec:survey} respectively, we move to the main goal of this paper, i.e. exploring the `systematic error budget' of our baseline 3x2 analyses as a function of survey progress of LSST (Y1, Y3, Y6, Y10) in section \ref{sec:science_return_map}. Specifically, we quantify how much cosmological constraints weaken when marginalizating over (combinations of) different systematic effects and when assuming different priors on the systematics parameters. In section \ref{sec:small_scales} we examine the gain in cosmological information when including small scales through improved baryonic and galaxy bias models. Finally, we explore how future spectroscopic surveys can improve cosmological constraints from LSST in section \ref{sec:spectro}  before concluding in section \ref{sec:summary}.


\section{Theoretical Basics}\label{sec:background}

\subsection{Theoretical modeling of the data vectors}\label{ssec:model_data_vector}

Our 3$\times$2  data vector is comprised of real-space two point correlation functions which requires the calculation of projected angular power spectra for the convergence field and galaxy density field. The angular power spectrum of two observables, $A$ and $B$ (here, $A/B$ stand for either the convergence field or the galaxy density field) for a Fourier mode, $\ell$ is given in terms of the 3D power spectrum $P_{AB}(k,z)$ as,

\begin{equation}\label{eqn:angular_power_spectra}
    C^{ij}_{AB}(\ell) = \int_0^{\chi_{\text{H}}} d\chi \frac{q^{i}_A(\chi) q^{j}_B(\chi)}{\chi^2} P_{AB}\bigg[ \frac{\ell + 1/2}{\chi}, z(\chi) \bigg],
\end{equation}
where, the Latin indices ($i/j$) label the tomographic bins. The weight kernels $q$ for the galaxy density field, $\delta_g$ and the convergence field, $\kappa$ are given as,
\begin{align}
    q^i_{\delta_g}(\chi) &= \frac{n^i[z(\chi)]}{\bar{n}^i} \frac{\diffop z}{\diffop \chi},\\
    q_{\kappa}(\chi) &= \frac{3 H^2_0 \Omega_m}{2c^2} \frac{\chi}{a(\chi)} \int_{\chi}^{\chi_{\text{H}}} \diffop \chi^{\prime} \frac{\diffop z}{\diffop \chi^{\prime}} \frac{n^i[z(\chi^{\prime})]}{\bar{n}^i} \frac{\chi^{\prime} - \chi}{\chi^{\prime}}.
\end{align}
In the above, $n^i(z)$ denotes the redshift distribution of galaxies in the $i$-th tomographic bin, $\chi$ denotes the comoving distance, $\chi_{\text{H}}$ denotes the comoving horizon distance, $a$ is the scale factor, $\bar{n}^i$ is the mean number density of source/lens galaxies. $H_0$ is the Hubble constant, $\Omega_m$ is the matter fraction of the Universe and $c$ is the speed of light. 

In case of the lens sample being different from the source sample, the $n(z)$ entering both equations are different, however we use the source sample itself as a lens sample; therefore we use the same $n(z)$ in both kernels. The power spectrum entering equation \eqref{eqn:angular_power_spectra} is related to the nonlinear matter power spectrum, $P_{\delta\delta}$, which we calculate assuming the {\sc halofit} prescription \citep{Takahashi2012}. The power spectra associated with the galaxy density field and the convergence field are given as,
\begin{align}
    P_{\delta_g B}(k,z) &= b_g(z) P_{\delta B}(k, z), \\
    P_{\kappa B}(k, z) &= P_{\delta B}(k, z),
\end{align}
where, $B$ stands for $\delta_g/\kappa$, and $b_g$ is the galaxy bias. In the above, we have also assumed the Limber approximation \citep{Limber1953, LoVerde2008}. 

The real space correlation functions, namely the cosmic shear $\xi_{\pm}$, galaxy-galaxy lensing, $\gamma_t$, and galaxy clustering, $w$,
are given in terms of the angular power spectra as \citep{Krause2021}, 
\begin{align}    
    \xi^{ij}_{\pm} &= \sum_{\ell} \frac{2\ell + 1}{2\pi \ell^2 (\ell + 1)^2}[G^{+}_{\ell,2}(\cos \theta) \pm G^{-}_{\ell,2}(\cos \theta)] C_{\kappa\kappa}^{ij}(\ell), \label{eqn:xi_cosmic_shear}\\
    \gamma^{ij}_t(\theta) &= \sum_{\ell} \frac{(2\ell + 1)}{2\pi \ell^2 (\ell + 1)^2} P^2_{\ell}(\cos \theta) C^{ij}_{\delta_g \kappa}, \label{eqn:gammat_ggl}\\
    w^{i}(\theta) &= \sum_{\ell} \frac{2\ell + 1}{4\pi} P_{\ell}(\cos \theta) C^{ii}_{\delta_g \delta_g} (\ell). \label{eqn:w_clustering}
\end{align}
In the above, $P_{\ell}$ and $P^2_{\ell}$ are the Legendre and associated Legendre polynomials. The functions $G^{\pm}_{\ell,2}$ are given in \citep{Stebbins1996}. Finally, to compare against the measurement of the correlation function, we average the correlation functions in angular bins. The details of the scale cuts and angular bins used in this work is given in section \ref{ssec:scale_cuts}.


\subsection{Simulated likelihood analysis}\label{ssec:simulated_analyses}

To gauge the impact of systematic effects on cosmological constraining power we conduct simulated likelihood analyses in a joint parameter space of cosmological parameters and systematics parameters.

We compute the posterior probability in parameter space as 
\begin{equation}\label{eqn:bayes}
    \mP(\mtheta|\md_{\text{sim}}) \propto \mL(\mtheta|\md_{\text{sim}}) \mP(\mtheta),
\end{equation}
where, $\mtheta$ is the collection of cosmological and systematic parameters and $\mP(\mtheta)$ is the prior on these parameters. In the above equation, $\mL$ is the likelihood function which is assumed to be Gaussian
\begin{equation}\label{eqn:log_lkl}
    \log \mL(\mtheta|\md_{\text{sim}}) = -\frac{1}{2} 
    [\md_{\text{model}}(\mtheta) - \md_{\text{sim}}]^{\text{T}} \mmat{C}^{-1} [\md_{\text{model}}(\mtheta) - \md_{\text{sim}}].
\end{equation}
The simulated data vectors $\md_{\text{sim}}$ are computed at a fiducial point in parameter space (see Table \ref{tbl:parameters}), the model vectors $\mvec{d}_{\text{model}}$ are computed as a function of cosmological and systematic parameters, and $\mmat{C}$ is the covariance of the data vector. 
In sections \ref{ssec:impact_studies} and \ref{sec:spectro}, we perform impact studies by computing contaminated $\mvec{d}_{\text{sim}}$ at different parameter values and then analyzing them with fiducial models. $\mvec{d}_{\text{sim}}$ are noiseless in the sense that no noise is added to the computed theory data vector.

We use the {\sc cocoa}\footnote{\url{https://github.com/CosmoLike/Cocoa}} framework \citep{Miranda:2020lpk} to compute the data and  model vectors following the prescription of the previous section. {\sc cocoa}({\sc cobaya}-{\sc cosmolike} {\sc architecture}) integrates \cosmolike \citep{Krause2017},  into the {\sc cobaya} likelihood framework \citep{Torrado2021}.

We use {\sc cosmocov} \citep{Fang2020} to calculate the covariance for our 3$\times$2  analyses. {\sc cosmocov}\footnote{\url{https://github.com/CosmoLike/CosmoCov}} is a publicly available code that uses analytical calculations to compute the covariance of the real-space correlation functions including the super-sample and the connected non-Gaussian part of the covariance. The code has been used to compute the 3$\times$2 real space covariances for the relevant Dark Energy Survey (DES) analyses \citep{DESY1, DES_3x2} and the Fourier covariance for the DESC-SRD \citep{DESC_SRD}. We refer the reader to \cite{Krause2017, Fang2020, Krause2021} for exact details of the covariance calculation.

In our cosmological analysis, we use the commonly used $w_0$-$w_a$ parameterization of the dark energy (DE) equation of state \cite{Chevallier2001, Linder2003}, where the redshift-dependent DE equation of state is given as,
\begin{equation}
    w(a) = w_0 + w_a \bigg(\frac{z}{1+z}\bigg).  
\end{equation}
In the later sections, we report our results in terms of $w_p$, which is the DE equation of state at a pivot redshift $z_p$, such that $w_0$ and $w_p$ are roughly uncorrelated. We use $z_p=0.4$ as our pivot redshift.

The calculation of the model vector during the MCMC is a significant computational cost. In this paper, the model vector is computed with the help of an emulator that speeds up the calculation by orders of magnitude (see section \ref{ssec:emulator}).


\subsection{Neural Network Emulator}\label{ssec:emulator}

In this paper we train the same 3 layer, fully connected neural network architecture (1024 neurons, ReLU activation function) as in \cite{Boruah2022} to predict the model data vector as a function of the cosmological and systematic parameters. We add one minor modification in that we train $4$ different neural networks to separately predict the $\xi_{+}$, $\xi_{-}$, $\gamma_t$ and $w$ parts of the model vector, instead of training one network for all parts. 

Even with extensive training, a brute force approach of building an emulator over the full prior range of the high-dimensional parameter space of LSST 3$\times$2 analyses did not yield sufficiently accurate results. Similar to \cite{Boruah2022} we solve this problem by combining two ideas: Firstly, we shrink the emulated parameter space by separating ``slow'' and ``fast'' parameters (also see \cite{Lewis2013} for this idea). The former are included in the emulator and the latter are applied after the emulator has returned its result. Examples for slow parameters are cosmological parameters, IA, and photo-$z$ parameters, examples for fast parameters in our implementation are baryonic uncertainties, galaxy bias, and shear calibration. Secondly, we implement an iterative design, where we acquire training samples from a tempered posterior that is obtained from an MCMC analysis using the emulator from the previous iteration. In that way, we acquire more training samples in the high posterior region, which is significantly smaller than the full prior range. The combination of these ideas gives excellent results for an LSST likelihood analysis after 3-4 iterations of training the neural network \citep[see][for details]{Boruah2022}. 

We note that this type of iterative local training reduces the walltime of an MCMC significantly since the calculation of the training sample can be fully parallelized during each iteration. Since the subsequent emulator based MCMC only takes minutes, the main computational bottleneck is the training of the neural network. Including emulator training, the walltime of such an MCMC is of order 1-2h; all subsequent chains using this trained emulator run within minutes, using one chain per core. 

\begin{figure*}
    \includegraphics[width=\linewidth]{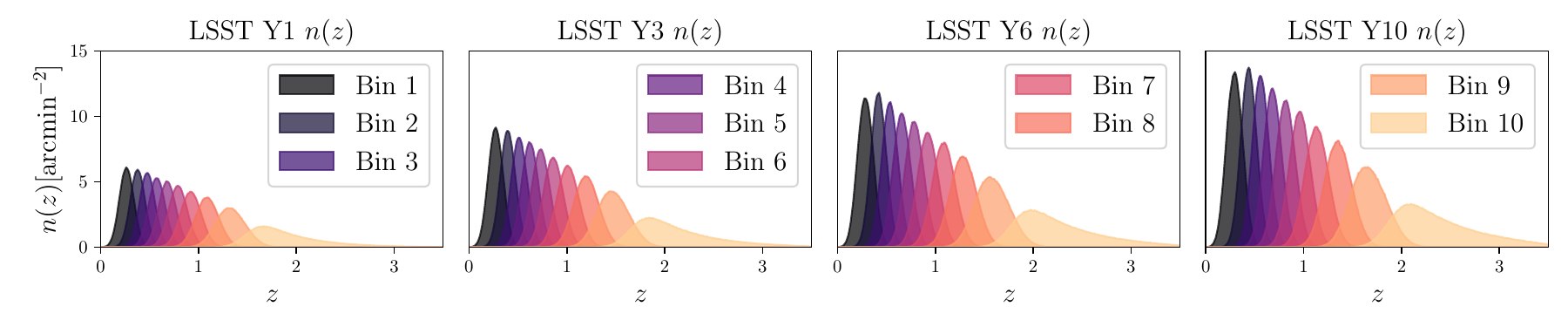}
    \caption{The redshift distribution, $n(z)$, for the LSST Y1/Y3/Y6/Y10 mock LSST surveys. The redshift distribution is generated according to the analytical distribution given in equation \eqref{eqn:smail} and then divided into $10$ tomographic bins with equal number of galaxies in each bin. The parameters $z_0, \alpha$ are given in Table \ref{tbl:survey_conf}.}
    \label{fig:lsst_nz}
\end{figure*}

Our locally trained emulator has limitations especially in the context of impact studies (see sections \ref{ssec:impact_studies}, \ref{ssec:photoz_outliers}), where the data vector is contaminated with a known systematic that is not perfectly captured with our default analysis pipeline (see Section \ref{ssec:impact_studies}). 

In that case the resulting posterior is offset from the region where the emulator was trained, which degrades its accuracy. We have explored this issue and find that the emulator is reliable if the best-fit parameter value of the contaminated data vector is within $\sim$ 2$\sigma$ of the fiducial cosmology and systematics parameter value. 

This accuracy is sufficient for most of the results presented in this paper. Where it is necessary (e.g, in Sections \ref{ssec:impact_studies}, \ref{ssec:photoz_outliers}, \ref{ssec:lowz_specz}), we retrain our emulator for the appropriate parameter regions. 

We emphasize the importance of exploring alternative neural network architectures that can cover a larger parameter space reliably, but postpone corresponding studies to future work.



\section{Survey assumptions and modeling choices}\label{sec:survey}

\begin{table}
  \centering
  \caption{Survey configurations for our mock LSST-Y1/Y3/Y6/Y10 surveys. $n_{\text{eff}}$ denotes the effective number density of the source sample. $z_0, \alpha$ are the parameters controlling the redshift distribution given in equation \ref{eqn:smail}. $\sigma_e$ and $\sigma_z$ denotes the shape noise and the photo-$z$ scatter respectively.}
  \begin{tabular}{l c c c c}
  \hline
     Parameter &  LSST-Y1 & LSST-Y3 & LSST-Y6 & LSST-Y10\\
    \hline
    Area & $12300$ & $12744$ & $13411$ & $14300$\\
    $n_{\text{eff}}$ & $10.47$ & $16.75$ & $22.52$ & $27.10$\\    
    $z_0$ & $0.193$ & $0.184$ & $0.179$ & $0.176$ \\
    $\alpha$ & $0.876$ & $0.828$ & $0.800$ & $0.783$ \\
    $\sigma_e$ & $0.26$ & $0.26$ & $0.26$ & $0.26$ \\
    $\sigma_z$ & $0.05$ & $0.05$ & $0.05$ & $0.05$ \\
    \hline
  \end{tabular}
  \label{tbl:survey_conf}
\end{table}

\begin{table*}
  \centering
  \caption{Fiducial value and the prior used on various cosmological and systematic parameters. In the table flat$[a,b]$ denotes a flat prior between $a$ and $b$, whereas Gauss$[\mu, \sigma^2]$ denotes a Gaussian prior with mean $\mu$ and standard deviation $\sigma$. See section \ref{ssec:systematics} for more detail on the modelling of the systematic parameters.}
  \begin{tabular}{l c c c c c}
  \hline
     Parameters &  Fiducial & LSST-Y1 Prior & LSST-Y3 Prior & LSST-Y6 Prior & LSST-Y10 Prior\\
    \hline
    \textbf{Cosmology} & & \\
    $\omega_\textrm{c}$ & $0.1274$ & flat$[0.01, 0.3]$ & flat$[0.01, 0.3]$ & flat$[0.05, 0.25]$ & flat$[0.05, 0.25]$\\
    $\log (A_\textrm{s} \times 10^{10})$ & $3.04452$ & flat$[2.5, 3.5]$ & flat$[2.5, 3.5]$ & flat$[2.65, 3.4]$ & flat$[2.65, 3.4]$\\
    $n_{\textrm{s}}$ & $0.97$ & flat$[0.87, 1.07]$ & 
    flat$[0.87, 1.07]$ & 
    flat$[0.91, 1.04]$ & 
    flat$[0.91, 1.04]$ \\
    $\omega_\textrm{b}$ & $0.0196$& flat$[0.01, 0.04]$ & flat$[0.01, 0.04]$ & flat$[0.01, 0.04]$ & flat$[0.01, 0.04]$\\
    $h$ & $0.70$& flat$[0.60, 0.80]$ & flat$[0.60, 0.80]$ & flat$[0.60, 0.80]$ & flat$[0.60, 0.80]$\\
    $w_0$ & $-1.$& flat$[-1.7, -0.3]$ & flat$[-1.7, -0.3]$ & flat$[-1.7, -0.3]$ & flat$[-1.7, -0.3]$\\
    $w_0 + w_a$ & $-1.$& flat$[-3.0, -0.1]$ & flat$[-3.0, -0.1]$ & flat$[-3.0, -0.1]$ & flat$[-3.0, -0.1]$\\
    \hline
    \textbf{Intrinsic Alignment} & & \\
    $a_{\ia}$ & $0.5$ & flat$[-5, 5]$ & flat$[-5, 5]$ & flat$[-5, 5]$ & flat$[-5, 5]$\\
    $\eta_{\ia}$ & $0$ & flat$[-5, 5]$ & flat$[-5, 5]$ & flat$[-5, 5]$ & flat$[-5, 5]$\\
    \hline
    \textbf{Linear galaxy bias} & & \\
    $b^{(i)}_1$&  & flat$[0.8, 3]$ & flat$[0.8, 3]$ & flat$[0.8, 3]$ & flat$[0.8, 3]$ \\
    \hline
    \textbf{Photo-$z$ bias} & & \\    $\Delta^i_{z,\text{source}}$ & 0 & Gauss$(0., 0.002^2)$ & Gauss$(0., 0.002^2)$ & Gauss$(0., 0.002^2)$ & Gauss$(0., 0.002^2)$\\
    \hline
    \textbf{Shear calibration} & & \\
    $m^i$ & 0 & Gauss$(0., 0.005^2)$ & Gauss$(0., 0.005^2)$ & Gauss$(0., 0.005^2)$ & Gauss$(0., 0.005^2)$\\
    \hline
    \textbf{Baryon PCA amplitude} & & \\
    $Q_1$ & 0 & flat$[-1000, 1000]$ & flat$[-1000, 1000]$ & flat$[-1000, 1000]$ & flat$[-1000, 1000]$\\
    $Q_2$ & 0 & flat$[-1000, 1000]$ & flat$[-1000, 1000]$ & flat$[-1000, 1000]$ & flat$[-1000, 1000]$\\
    $Q_{3}$ & 0 & flat$[-1000, 1000]$ & flat$[-1000, 1000]$ & flat$[-1000, 1000]$ & flat$[-1000, 1000]$\\
    \hline
  \end{tabular}
  \label{tbl:parameters}
\end{table*}

\subsection{LSST survey assumptions}\label{ssec:lsst_assumptions}

We closely follow the LSST-DESC SRD \citep{DESC_SRD} to design our Y1, Y3, Y6, and Y10 mock surveys. As specified in the DESC-SRD, we assume the Y1 survey area to be 12300 deg$^2$ and the Y10 survey area to be 14300 deg$^2$. We interpolate the areas of Y3 and Y6 linearly between Y1 and Y10, resulting in survey areas of 12744 deg$^2$ and 13411 deg$^2$ for Y3 and Y6, respectively. We note that survey optimization is an ongoing effort in LSST and that the final footprints will likely look differently. 

In this paper, we consider a ``lens=source" analysis \citep[see e.g.][]{Fang2022} using the source sample as defined in the DESC-SRD also as the lens sample.

The resulting redshift distribution is fit to the parametric form  \citep{Smail1995},
\begin{equation}\label{eqn:smail}
    n(z) \propto z^2 \exp[-(z/z_0)^{\alpha}].
\end{equation}
The effective number density and the values of $z_0$ and $\alpha$ are obtained as a function of $i_{\text{depth}}$ which can be approximated using the following analytic formulae,

\begin{align}
    n_{\text{eff}} &= 10.47 \times 10^{0.3167 (i_{\text{depth}} - 25)} \\
    z_0 &= 0.193 - 0.0125 (i_{\text{depth}} - 25) \\
    \alpha &= 0.876 - 0.069 (i_{\text{depth}} - 25).
\end{align}
We assume that the LSST-Y1, Y3, Y6 and Y10 surveys will have $i_{\text{depth}} = 25.1, 25.7, 26.1, 26.35$ respectively. The resulting redshift distributions are convolved with a Gaussian photo-$z$ error of standard deviation, $\sigma_z = 0.05(1+z)$. The galaxies are then binned into $10$ tomographic bins of equal galaxy number density. The resulting redshift distributions are shown in Figure \ref{fig:lsst_nz} with effective number density for LSST-Y1/Y3/Y6 and Y10 being $n_{\text{eff}} = 10.47, 16.75, 22.52$ and $27.1$ arcmin$^{-2}$, respectively (also see Table \ref{tbl:survey_conf}). In section \ref{ssec:photoz_outliers}, we will also consider the impact of photo-$z$ outliers, which changes the redshift distribution of the galaxies.

\subsection{Systematics modelling}\label{ssec:systematics}

Our analysis includes observational systematic effects such as photo-$z$ uncertainties, shear biases and astrophysical systematics including baryons, intrinsic alignment and galaxy bias. 

\subsubsection{Photo-$z$}\label{sssec:photoz}

Inaccuracies in the photometric redshift distribution, $n^i(z)$ are modeled as a simple shift model that adds a new parameter, $\Delta^i_z$, to each tomographic bin, which shifts the redshift distribution as,
\begin{equation}
    n^i(z) \rightarrow n^i(z + \Delta^i_z). 
\end{equation} 
In ``lens=source" analyses, we use the same $\Delta_z$ parameter for both the lens and source samples. This reduces the dimensionality of the parameter space, compared to using a different lens sample, and results in improved constraints on photometric redshift parameters and cosmological parameters \citep{Fang2022}. Our baseline analysis follows the DESC-SRD and assumes that the redshift distributions can be characterized with a precision of $0.2\%$ ($\sigma[\Delta_z] = 0.002$). In section \ref{ssec:priors}, we examine the impact of a more pessimistic photo-$z$ prior. Photo-$z$ parameters are slow parameters that are included in the emulator training.

\subsubsection{Baryons}\label{sssec:baryons}
Baryons impact matter distribution on small scales, which in turn affects the 3$\times$2 observables. We use the formalism of \cite{Eifler2015, Huang2019} to account for the effects of baryons. In this formalism, we measure the changes in the matter power spectrum caused by baryonic physics in various hydrodynamical simulations.  In this paper, we use $8$ hydrodynamical simulations: Illustris TNG-100 \citep{Springel2018}, MassiveBlack-II \citep{Khandai2015}, Horizon-AGN \citep{Dubois2014}, 2 simulations (T8.0 and T8.5) of the cosmo-OWLS simulation suite \citep{LeBrun2014} and 3 simulations (T7.6, T7.8, T8.0) of the BAHAMAS simulation suite \citep{McCarthy2017}. 

We perform a Principal Component Analysis (PCA) to capture said difference as it affects our model vectors (equations  \eqref{eqn:xi_cosmic_shear}-\eqref{eqn:w_clustering}) with only a few modes. 
The 3$\times$2 data vector including baryonic physics is then modeled in terms of the principal components $\text{PC}_i$ as,
\begin{equation}
    \mvec{d}_{\text{baryons}} = \mvec{d}_{\text{dmo}} + \sum_{i=1}^{N_{\text{PCA}}} Q_i \text{PC}_i,
\end{equation}
where $\{Q_i\}$, the amplitude of the $\text{PC}_i$, represent the additional nuisance parameters varied in a likelihood analysis and $\mvec{d}_{\text{dmo}}$ is the dark matter only data vector. In this paper, we use $N_{\text{PCA}} = 3$ following \cite{Huang2019} who explored the number of PCs necessary for an LSST Y10 analysis. The $\{Q_i\}$ are fast parameters, hence not included in the emulator training.

\subsubsection{Intrinsic alignment}\label{sssec:ia}

We model the intrinsic alignment using a redshift-dependent nonlinear alignment (NLA) model \citep{Bridle2007}. In this model, the change in the angular power spectrum due to IA is modelled using a redshift dependent amplitude,
\begin{equation}
    A_{\text{IA}}(z) = -a_{\ia}\frac{C_1 \rho_{\text{cr}} \Omega_m}{G(z)}\bigg(\frac{1+z}{1+z_{0,\text{IA}}}\bigg)^{\eta_{\ia}},
\end{equation}
where, $C_1 \rho_{\text{cr}} = 0.0134$, $z_{0,\text{IA}} = 0.62$, $G(z)$ is the linear growth factor. There are only two free parameters in this model -- $a_{\ia}, \eta_{\ia}$ that parameterize the amplitude and the redshift dependence of the IA signal, respectively. For further details on our exact intrinsic alignment modeling, we refer to \cite{Krause2016, DESC_SRD, Krause2021}. 

For additional complexity of IA modeling we refer to \cite{Blazek2019,Vlah2020,Fortuna2021}. IA parameters are slow parameters that are included in the emulator training. 

\begin{figure*}
    \centering
    \includegraphics[width=\linewidth]{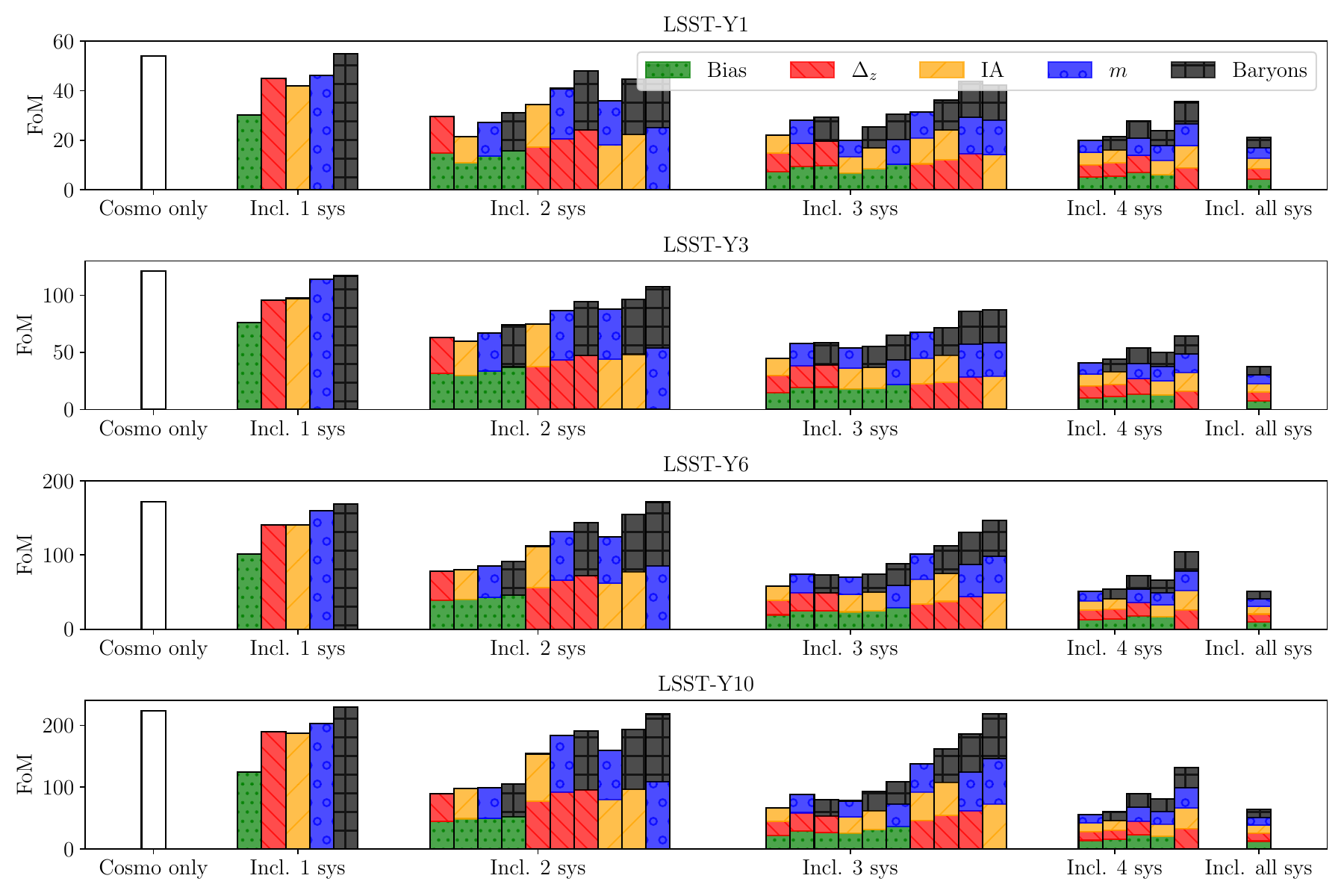}
    \caption{$w_0$-$w_a$ Figure-of-merit budget for all combinations of systematics. The different colors in the legend shows systematic parameter that is varied in a given chain. For example, when the red bar is present, the photo-$z$ bias parameters are varied. The size of the different bars are equal and \textit{do not} denote the contribution of different systematics. When multiple systematic parameter combinations are varied, it captures the contribution from the correlated systematic parameters in the chain.}
    \label{fig:fom_combinations}
\end{figure*}

\subsubsection{Shear multiplicative bias}\label{sssec:shear_m}

We model possible biases in the shear inference through a multiplicative factor
\begin{equation}
    \gamma_{\text{measured}} = (1 + m)\times \gamma_{\text{true}}.
\end{equation}
Specifically, we add an extra parameter $m^i$ for each tomographic bin, which modifies the modeled observables as:
\begin{align}
    \xi^{ij}_{\pm} &\rightarrow (1 + m^i) (1 + m^j) \xi^{ij}_{\pm}, \\
    \gamma^{ij}_{t} &\rightarrow (1 + m^i) \gamma^{ij}_{t}.
\end{align}
The $m^i$ are fast parameters, hence not included in the emulator training.

\subsubsection{Galaxy bias}\label{sssec:gbias}
Following the DESC SRD we only include linear galaxy bias in our model. As we will see in section \ref{ssec:impact_studies}, this choice does not bias cosmological constraints given the DESC SRD scale cuts. The $b^{i}_g$ are fast parameters, hence not included in the emulator training. The training data vectors are computed at the fiducial galaxy bias, $b^{i}_{g,\text{fid}}$.
Linear galaxy bias then changes the galaxy-galaxy lensing and the galaxy clustering part of the data vector as,
\begin{align}
    \gamma^{ij}_{t}(b^{j}_g) &= \bigg( \frac{b^{j}_g}{b^{j}_{g,\text{fid}}} \bigg) \times \gamma^{ij}_{t}(b^{j}_g = b^{j}_{g,\text{fid}}), \\
    w^i(b^{i}_g) &= \bigg(\frac{b^{i}_g}{b^{i}_{g,\text{fid}}}\bigg)^2 \times w^i (b^{i}_g = b^{i}_{g,\text{fid}}). \label{eqn:galaxy_clustering_scaling}
\end{align}
The fiducial value for the galaxy bias in the $i$-th redshift bin is set to $b_i = 1.05 / G(\bar{z}_i)$, where $G$ is the growth function and $\bar{z}_i$ is the mean redshift of the $i$-th redshift bin \cite{Nicola2020}. 

We compute the galaxy clustering part of the 3$\times$2 pt data vector using the  non-Limber prescription of \cite{Fang2020nonlimber}, but neglecting contributions from magnification bias. In addition to the density-density terms, galaxy bias also enters through cross-terms of density and redshift-space distortion. We opted to only vary galaxy bias as a fast parameter for the first term and neglect variations of the latter. We have confirmed that this approximation does not impact our contours.  

The fiducial values and the prior used for the various cosmological and systematics parameters are given in Table \ref{tbl:parameters}.

\subsection{Angular binning and scale cuts}\label{ssec:scale_cuts}

We bin the data vectors in 20 logarithmically spaced angular bins, ranging from 1 arcminute to 400 arcminutes. To minimize the impact of baryonic physics and nonlinear galaxy bias on cosmological results we adopt the DESC-SRD scale cuts with minor changes given that our 3$\times$2 analysis is conducted in real space. The scale cuts for galaxy-galaxy lensing and galaxy clustering are obtained by applying the real space cut of $21~h^{-1}$ Mpc (approximately corresponding to the DESC-SRD cut of $k_{\text{max}} = 0.3~h$ Mpc$^{-1}$) to angular scales. For cosmic shear the DESC-SRD's $\ell_{\text{max}} = 3000$ limit is converted to real-space as $\theta_{\text{min}} = 2.756 (8.696)$ arcmin for $\xi_{+/-}$, so that $\ell_{\text{max}} \theta_{\text{min}}$ corresponds to the first zero of the Bessel function $J_{0/4}$. 

In section \ref{sec:small_scales}, we evaluate the science return and potential biases from imperfect modeling choices by extending the analysis to smaller scales.



\begin{figure*}
    \centering
    \includegraphics[width=0.24\linewidth]{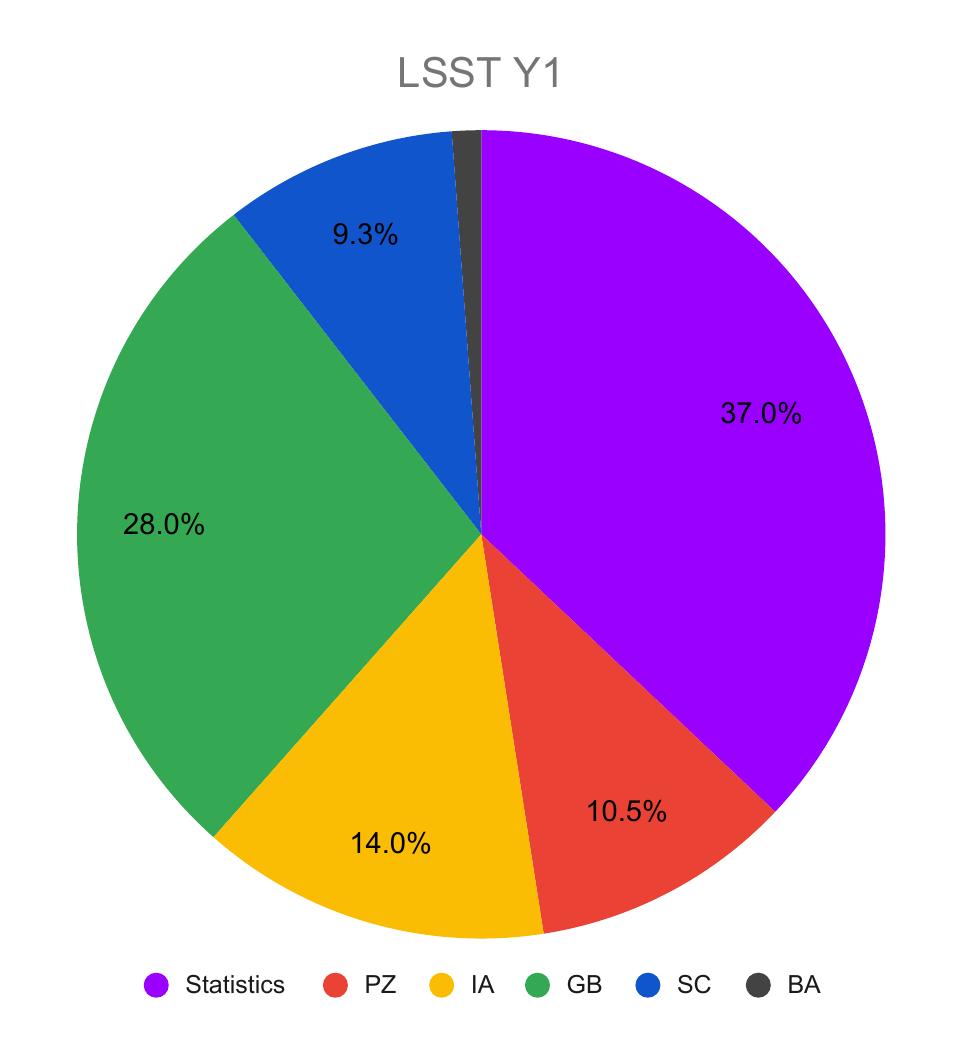}
     \includegraphics[width=0.24\linewidth]{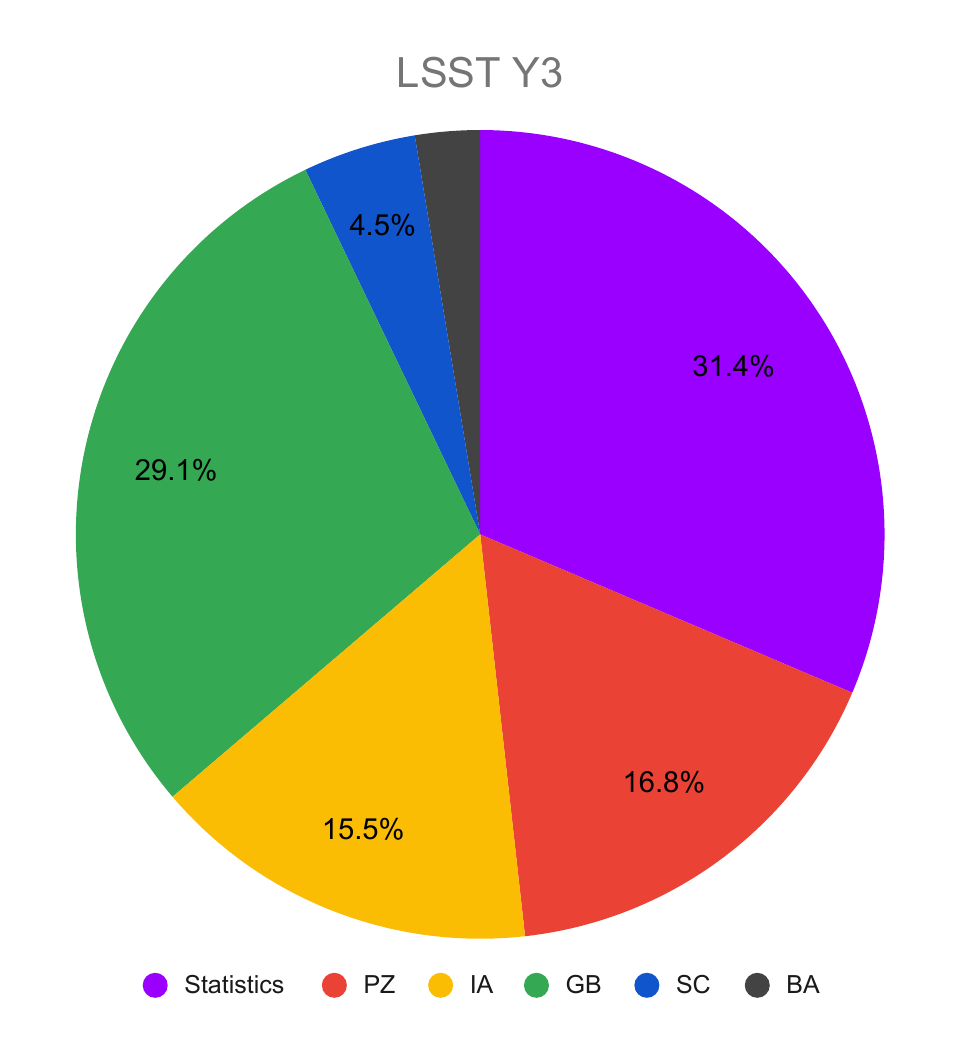}
      \includegraphics[width=0.24\linewidth]{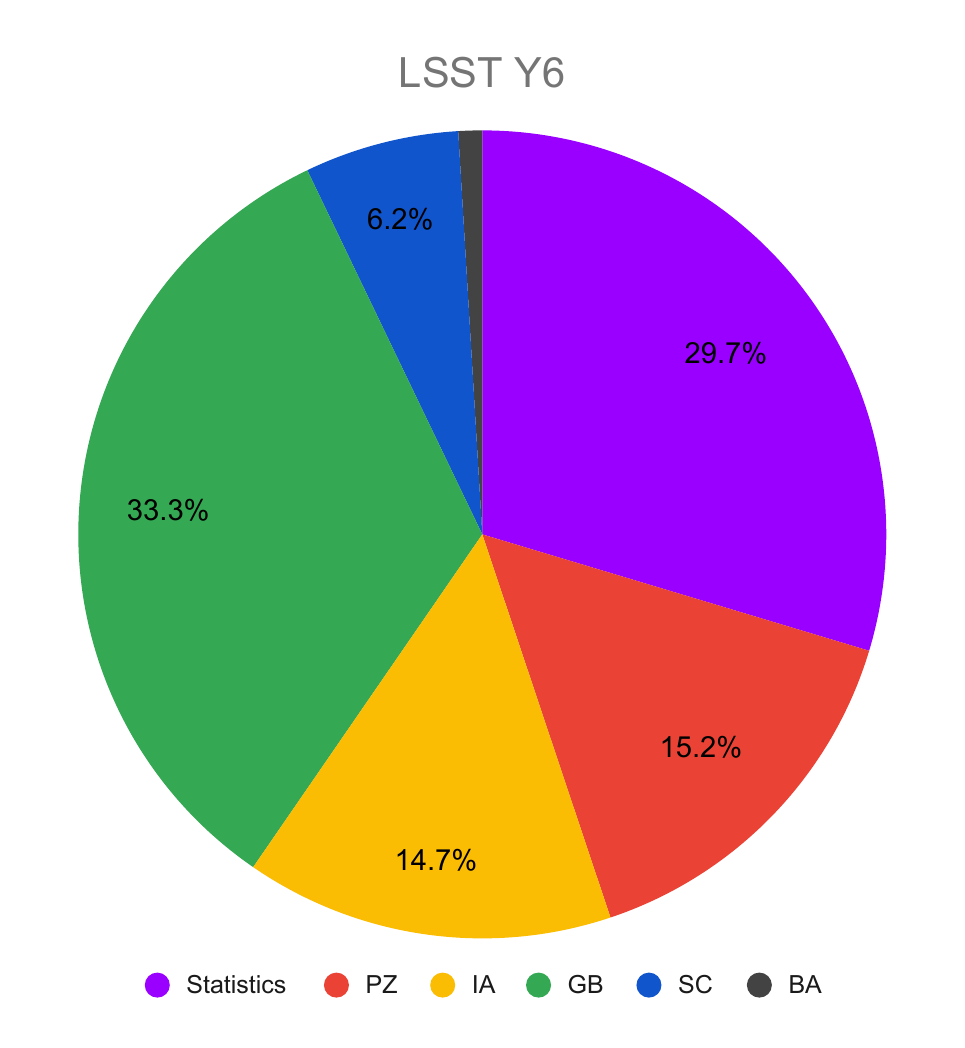}
       \includegraphics[width=0.24\linewidth]{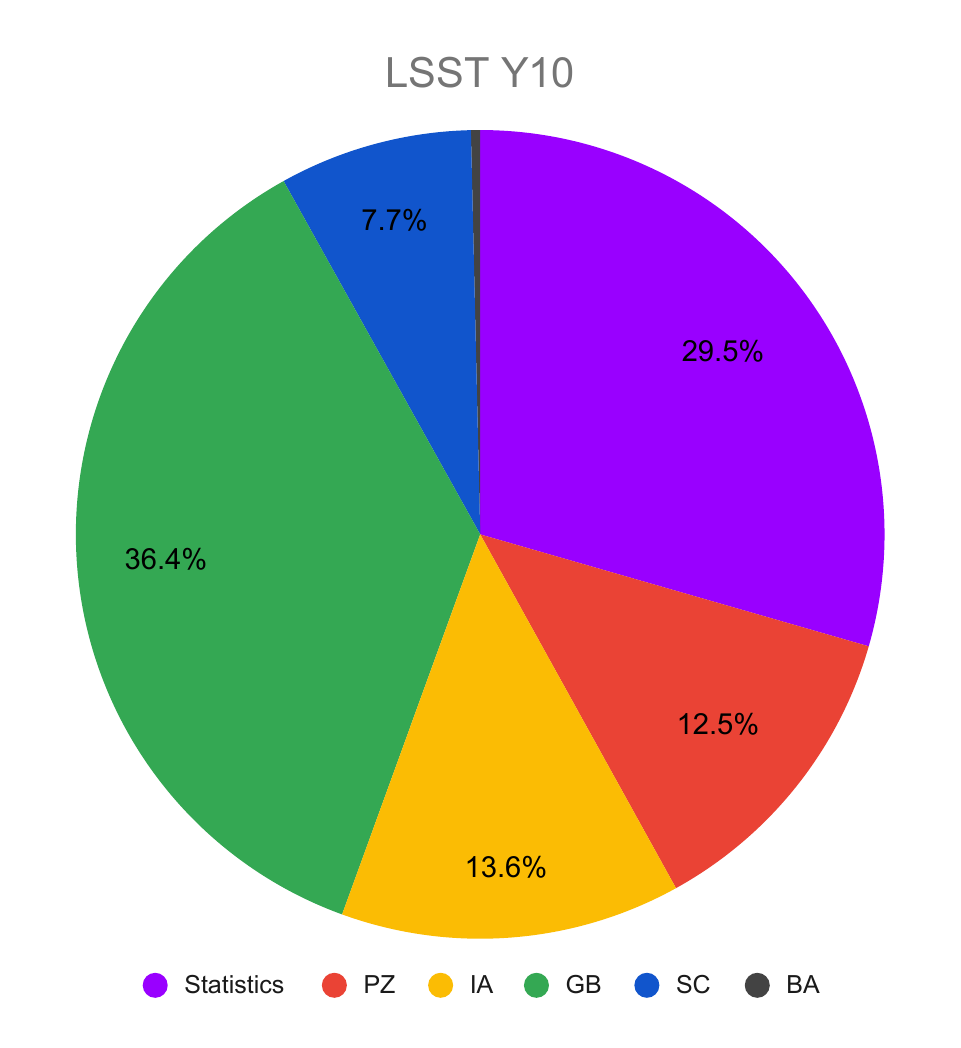}
    \caption{Fractional uncertainties quantified through the dark energy figure of merit based on Table \ref{tbl:sys_error_budget} (see text for calculation details). We stress that these values change depending on the very specific analysis choices (scale cuts, systematics parameterization, parameter priors) assumed. However, for this specific analysis, the takeaway is that the wide ranging prior on galaxy bias is the largest source of systematic uncertainty, followed by IA and PZ, then shear calibration. Tighter priors on our baryon PC amplitudes will only be useful for different analysis choices, e.g. different scale cuts. }
    \label{fig:systematics_error_budget}
\end{figure*}

\begin{table*}
  \centering
  \caption{The dark energy figure-of-merit for different combination of systematic parameters. }
  \begin{tabular}{l c c c c}
  \hline
     Varied systematic parameters &  LSST-Y1 & LSST-Y3 & LSST-Y6 & LSST-Y10\\
    \hline
    \textbf{None (Cosmo only)} & 54 & 121 & 172 & 224 \\
    \hline
    \textbf{Vary 1 set of systematic parameters} & & & & \\
    Photo-$z$ & 45 & 95 & 140 & 190 \\    
    IA & 42 & 97 & 141 & 187 \\    
    Galaxy bias & 30 & 76 & 102 & 125 \\    
    Shear calibration & 46 & 114 & 159 & 203 \\    
    Baryons & 53 & 117 & 170 & 223 \\    
    \hline
    \textbf{Vary 2 sets of systematic parameters} & & & & \\
    Photo-$z$, IA & 34 & 75 & 112 & 154 \\  
    Photo-$z$, Galaxy bias & 29 & 63 & 78 & 89 \\  
    Photo-$z$, Shear calibration & 41 & 86 & 132 & 184\\  
    Photo-$z$, Baryons & 45 & 94 & 139 & 191 \\  
	IA, Galaxy bias & 22 & 60 & 80 & 98  \\  
    IA, Shear calibration & 36 & 88 & 124 & 160  \\  
    IA, Baryons & 42 & 96 & 142 & 193 \\  
    Galaxy bias, Shear calibration & 27 & 67 & 85 & 99\\  
    Galaxy bias, Baryons & 30 & 74 & 92 & 105 \\  
    Shear calibration, Baryons & 46 & 107 & 160 & 218 \\  
    \hline
    \textbf{Vary 3 sets of systematic parameters} & & & & \\
    Galaxy bias, Shear calibration, Baryons & 30 & 65 & 88 & 109 \\  
    IA, Shear calibration, Baryons & 42 & 87 & 145 & 219 \\  
    IA, Galaxy bias, Baryons & 25 & 55 & 74 & 93 \\  
    IA, Galaxy bias, Shear calibration, & 20 & 54 & 70 & 78 \\  
	Photo-$z$, Shear calibration, Baryons & 44 & 86 & 131 & 186\\  
    Photo-$z$, Galaxy bias, Baryons & 29 & 59 & 73 & 80 \\  
    Photo-$z$, Galaxy bias, Shear calibration & 28 & 58 & 74 & 88 \\  
    Photo-$z$, IA, Baryons & 36 & 71 & 112 & 162 \\  
    Photo-$z$, IA, Shear calibration & 31 & 67 & 101 & 138 \\  
    Photo-$z$, IA, Galaxy bias & 22 & 45 & 58 & 67 \\  
    \hline
    \textbf{Vary 4 sets of systematic parameters} & & & & \\    
	IA, Galaxy bias, Shear calibration, Baryons [Photo-$z$ fixed] & 24 & 50 & 66 & 81 \\  
    Photo-$z$, Galaxy bias, Shear calibration, Baryons [IA fixed] & 28 & 54 & 72 & 90 \\  
    Photo-$z$, IA, Shear calibration, Baryons [Galaxy bias fixed] & 35  & 66 & 104 & 152 \\  
    Photo-$z$, IA, Galaxy bias, Baryons [Shear calibration fixed] & 21 & 44 & 54 & 64 \\  
	Photo-$z$, IA, Galaxy bias, Shear calibration [Baryons fixed] & 20 & 41 & 51 & 61 \\  
    \hline
    \textbf{Vary all systematics} & 20 & 38 & 51 & 61 \\  
    \hline
  \end{tabular}
  \label{tbl:sys_error_budget}
\end{table*}

\section{Mapping impact of systematics}\label{sec:science_return_map}

In this paper we pursue two different avenues to quantify the impact of systematics: Firstly, we consider the increase in uncertainty when marginalizing over nuisance parameters associated with the different types of systematics. Secondly, we study parameter biases that occur when not (properly) accounting for systematics in the analyses. We note that both approaches are affected by the exact parameterization that is assumed for the systematics and results will change as a function of said parameterization and also as a function of other analysis choices, in particular scale cuts (see section \ref{sec:small_scales}) and priors (see section \ref{ssec:priors}).

\subsection{Systematics error budget}\label{ssec:error_budget}

As described in section \ref{ssec:systematics}, we marginalize over 5 different groups of systematic effects, namely, photo-$z$ bias, intrinsic alignment, galaxy bias, shear multiplicative bias, and baryonic modeling uncertainties. We quantify the contribution of each of these systematic sets to the overall error budget individually and in combination. 

In Figure \ref{fig:fom_combinations} we show 128 simulated 3$\times$2  MCMC analyses and quantify the science return with the standard dark energy figure-of-merit (see Table \ref{tbl:sys_error_budget} for the exact numerical values). Each row corresponds to a different LSST dataset (from top to bottom: Y1, Y3, Y6, Y10). We increase the complexity of the analysis (the size of the parameter space) from left to right, starting with the cosmology-only analysis and ending with the full likelihood analysis that contains all $5$ groups of systematic effects.

Comparing the cosmology-only analysis on the far left column to the next set, where we marginalize over each group of the systematics individually, we clearly see that the largest degradation in constraining power comes from the inclusion of galaxy bias uncertainties (green). We further find that intrinsic alignment (yellow) and photo-$z$ uncertainties (red) also cause a significant loss in constraining power, with multiplicative shear calibration (blue) being less important, and baryonic uncertainties (dark gray) barely degrading the information content. 

This is further illustrated in Figure \ref{fig:systematics_error_budget}, which quantifies the error budget contributions from statistical uncertainties and the various systematics. Based on Table \ref{tbl:sys_error_budget}, we calculated the systematics budget as the difference between the figures of merit in the first and last rows of the table. We then further subdivide the systematics budget based on the relative degradation of the ``Cosmo only'' FOM values and the FOMs in the bracket below where we vary the systematics sets individually.

While some details vary for the Y1/Y3/Y6/Y10 scenarios, the lack of more stringent priors on galaxy bias parameters or a less flexible parameterization of galaxy bias is the largest contributor to the systematic error budget by far and consequently poses the largest opportunity to improve. Intrinsic alignment and photo-$z$ uncertainties follow as the next leading concerns and finally shear calibration and baryonic physics uncertainties contribute least to the error budget. 


Our analysis above provides a ranking of how much the dark energy equation of state parameters can be improved by finding better priors for the different systematics groups. However, it is important to note that this ranking will likely look different for other cosmological parameters of interest, e.g. $S_8$ or $H_0$. Further, this ranking can also look different if a different parameter space is considered, e.g. more complex dark energy or modified gravity models, or when different scale cuts are employed. Each parameter space is unique and we stress that a study as the one conducted in Figure \ref{fig:fom_combinations} should be performed early on when defining the analysis choices for a given science case. Our neural network emulator allows for such studies at scale. 

The ranking expressed by Figure \ref{fig:fom_combinations} and Table \ref{tbl:sys_error_budget} will also change if the parameterization of the groups of systematics changes. For example, a parameterization of baryonic physics uncertainties that has more degrees of freedom will impact the cosmological parameter space differently and can also cause a stronger degradation for dark energy parameters. Vice versa, a more confined parameterization of galaxy bias can reduce the degradation we see above. Lastly, we note that galaxy bias and baryonic physics parameterizations are inherently coupled to the question of scale cuts, which we explore further in Section \ref{sec:small_scales}.

For the exact analysis choices of this paper as described in Section \ref{sec:background} we find that LSST 3$\times$2 point analysis are systematics limited in the sense that systematic effects will contribute more to the overall error budget than statistical uncertainties. Comparing the cosmology-only FoMs in Table \ref{tbl:sys_error_budget} with the FoM where all systematic parameters are varied, we find a factor of 2-3 difference depending on the dataset.

\begin{figure*}
    \centering
    \includegraphics[width=\linewidth]{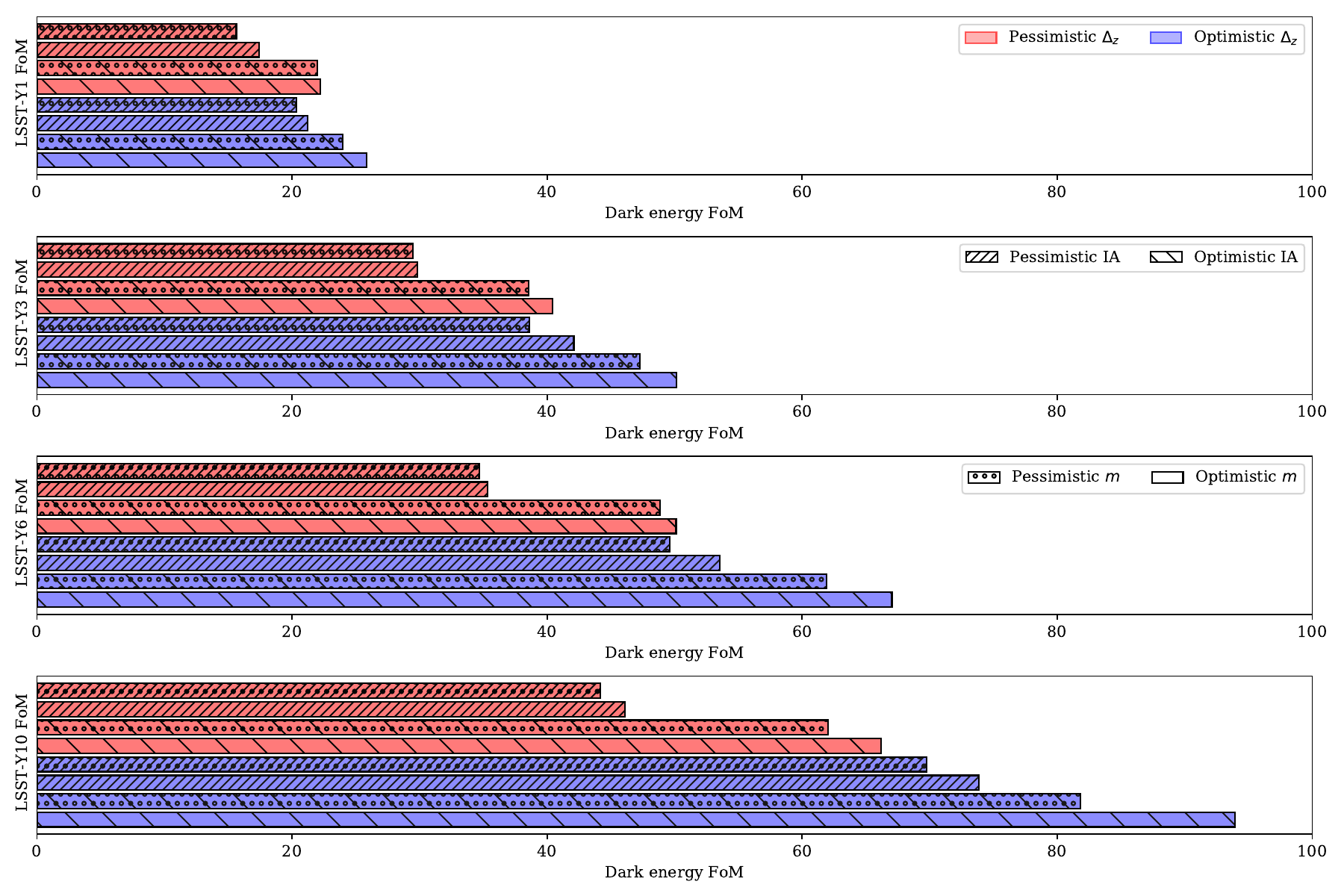}
    \caption{Dark energy Figure of merits for different combinations of optimistic and pessimistic priors. The blue (red) bars correspond to optimistic (pessimistic) prior on $\Delta_z$. The optimistic and pessimistic priors on the intrinsic alignment are denoted with a backslash (\textbackslash) and a dense right slanted line hatch ($/$) hatch respectively. The pessimistic priors on shear calibration bias is denoted with dots and the optimistic prior is left unfilled. The corresponding values are given in Table \ref{tbl:priors}.}
    \label{fig:priors_plot}
\end{figure*}

\subsection{Information gain from improved priors}\label{ssec:priors}

\begin{table}
  \centering
  \caption{Optimistic and pessimistic priors used in section \ref{ssec:priors}}
  \begin{tabular}{l c c c}
  \hline
     Systematic &  Pessimistic prior & Optimistic prior\\
    \hline
    \textbf{IA} &  & \\
    $a_1$ & Flat[-5,5]& Gauss[$0.5$, $0.019^2$] for LSST-Y1\\
		 &  & Gauss[$0.5$, $0.015^2$] for LSST-Y3\\
		 &  & Gauss[$0.5$, $0.013^2$] for LSST-Y6\\
		 &  & Gauss[$0.5$, $0.012^2$] for LSST-Y10\\
    $\eta_1$ & Flat[-5,5]& Gauss[$0$, $0.145^2$] for LSST-Y1\\
	& & Gauss[$0$, $0.110^2$] for LSST-Y3\\
	& & Gauss[$0$, $0.091^2$] for LSST-Y6\\
	& & Gauss[$0$, $0.079^2$] for LSST-Y10\\
    \hline 
    \textbf{Photo-$z$ bias} & & \\
    $\Delta_z$ & Gauss(0,$0.005^2$) & Gauss(0,$0.002^2$) for LSST-Y1\\    
    & & Gauss(0,$0.0015^2$) for LSST-Y3\\  
    & & Gauss(0,$0.0015^2$) for LSST-Y6\\  
    & & Gauss(0,$0.001^2$) for LSST-Y10\\  
    \hline   
    \textbf{Shear bias} & & \\
   $m$ & Gauss(0,$0.01^2$) & Gauss(0,$0.005^2$) \\
    \hline 
  \end{tabular}
  \label{tbl:priors}
\end{table}

\begin{table}
  \centering
  \caption{Dark energy figure-of-merit for different combinations of optimistic and pessimistic (shortened as `Opti' and `Pessi' in the table)
  priors defined in Table \ref{tbl:priors}. The results are shown in Figure \ref{fig:priors_plot}.}
	\begin{tabular}{l|c|c|c c c c}
\hline
$\Delta_z$ prior & IA prior & $m$ prior & Y1 & Y3 & Y6 & Y10\\
\hline
\multirow{4}{*}{Pessi} & \multirow{2}{*}{Pessi} & Pessi & 16 & 29 & 35 & 44 \\\cline{3-7}
& & Opti & 17 & 30  & 36  & 46 \\\cline{2-7}
    & \multirow{2}{*}{Opti} & Pessi & 22 & 39 & 48 & 62  \\\cline{3-7}
  & & Opti & 22 & 40 & 50 & 66 \\\hline
\multirow{4}{*}{Opti} & \multirow{2}{*}{Pessi} & 
Pessi & 20 & 39 & 50 & 70 \\\cline{3-7}
    & & Opti & 21  & 42  & 54  & 74 \\\cline{2-7}
    & \multirow{2}{*}{Opti} & Pessi & 24 & 47 & 62 & 82 \\\cline{3-7}
    &  & Opti & 26 & 50  & 67 & 94  \\\hline
\end{tabular}
\label{tbl:sys_priors}
\end{table}

In this section we explore the gain in constraining power when improving the prior knowledge on systematics parameters. These gains can be achieved via external observations, simulations, or algorithmic development. For example, future spectroscopic surveys in the LSST footprint \citep{Astro2020_spectroscopy1, Astro2020_spectroscopy2, Snowmass, megamapper} can have significant impact on photo-$z$ \cite{Masters2017, Newman2022, McCullough2023} and intrinsic alignment uncertainties \citep{Singh2015, Johnston2019, Samuroff2022}. We simulate different scenarios by imposing optimistic and pessimistic priors on different systematic parameters as shown in Table \ref{tbl:priors} and summarize our results in Figure \ref{fig:priors_plot} and Table \ref{tbl:sys_priors}.

\begin{itemize}
    \item \textbf{Intrinsic alignment}: 
    \cite{Astro2020_spectroscopy1} demonstrated that proposed wide-field spectroscopy of LSST galaxies can yield a 2$\times$ improvement in the signal-to-noise of the IA measurements. 
    In this paper we chose our fiducial priors on $a_1$ and $\eta_1$ (see Table \ref{tbl:priors}) as a pessimistic scenario and 
    a Gaussian prior with a standard deviation half the uncertainty of the fiducial analysis as the optimistic scenario.
    
    \item \textbf{Photo-$z$ bias}: Photo-$z$ uncertainties can be improved using larger spectroscopic samples \citep{Astro2020_spectroscopy1, Astro2020_spectroscopy2, Masters2017, McCullough2023} or better algorithms \citep{Sanchez2019,Wright2020,Alsing2023,Leistedt2023}. 
    For the optimistic scenario, we assume that the photo-$z$ biases can be constrained at the level of LSST requirements \citep{DESC_SRD}, $\sigma[\Delta_z] = 0.002 (0.001)$ for LSST-Y1 (Y10). The optimistic prior for the LSST-Y3/Y6 is taken to be, $\sigma[\Delta_z] = 0.0015$, an intermediate values between the LSST-Y1 and Y10 optimistic values.     
    Our pessimistic scenario assumes that this accuracy will be hard to achieve \citep{Newman2022} and  that the photo-$z$'s can be characterized with a standard deviation of $\sigma[\Delta_z] = 0.005$ only.
    
    \item \textbf{Shear multiplicative bias}: We follow the DESC-SRD to assume a shear multiplicative bias of $\sigma_m = 0.005$ in the optimistic case. As a pessimistic scenario, we assume a standard deviation of $\sigma_m = 0.01$.
\end{itemize}

We run our MCMC chains with all combinations of the optimistic and pessimistic priors on the 3 groups of systematics considered: intrinsic alignment, photo-$z$ biases and shear multiplicative bias. We do not consider uncertainties due to baryonic physics and galaxy bias in this section, since the parameterization of these small-scale systematics is a topic of active research and strongly dependent on scale cuts.

The results of these analyses are presented in Table \ref{tbl:sys_priors} and Figure \ref{fig:priors_plot}. 
As we can see, shear multiplicative bias does not impact the results even if we are only able to determine it at a $\sim 1\%$ level (twice the value of DESC-SRD). On the other hand, strongly constraining IA and photo-$z$ biases can lead to stronger constraints on cosmological parameters.  

For the pessimistic scenario in $\Delta_z$, the dark energy figure-of-merit is degraded by $\sim 20\%$ (for Y1) to $\sim 40\%$ (for Y10). Similarly, constraining IA better by a factor of 2 (optimistic scenario for IA) leads to an improvement in FoM of $\sim 30$-$40\%$. When considering the optimistic scenario for both IA and photo-$z$, this leads to an improvement of over 50$\%$ for LSST-Y1 and almost $100\%$ improvement for LSST-Y10. 



\begin{figure*}
    \centering
    \includegraphics[width=\linewidth]{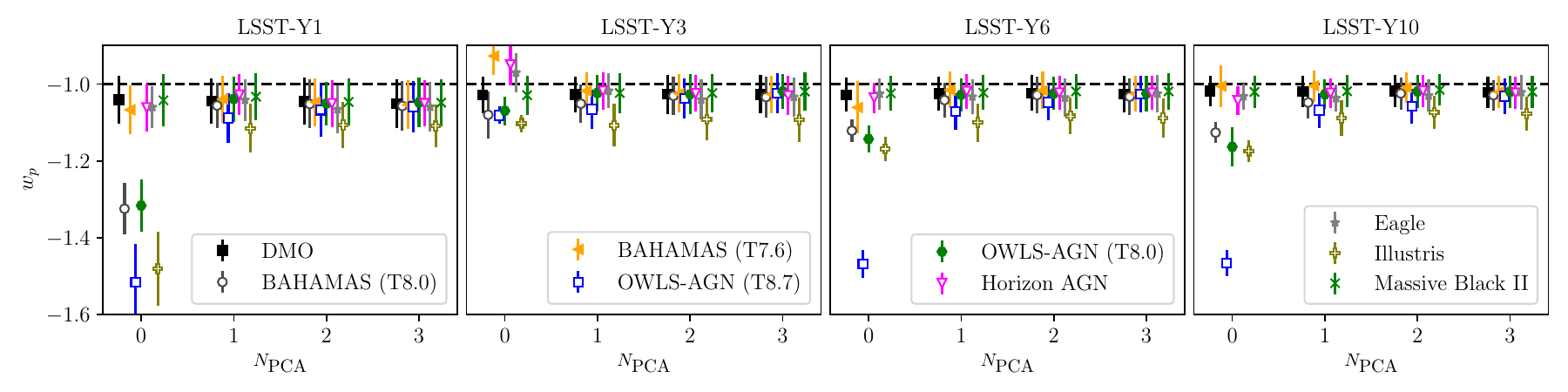}
    \caption{Impact of baryonic feedback on $w_p$ for LSST Y1/Y3/Y6 and Y10 mock surveys. We plot the marginalized 1D posteriors of $w_p$ for our analysis of the contaminated data vectors with varying number of baryon PCA modes used to account for baryons. The true parameter value is denoted with the black dashed lines. As we can see, without marginalizing over the baryon feedback parameters, the resulting cosmological parameters are heavily biased. However, marginalizing over 1 to 2 PC modes is sufficient to capture the impact of most small-scale baryon feedback scenarios.}
    \label{fig:baryon_impact_study}
\end{figure*}

\begin{figure*}
    \centering
    \includegraphics[width=\linewidth]{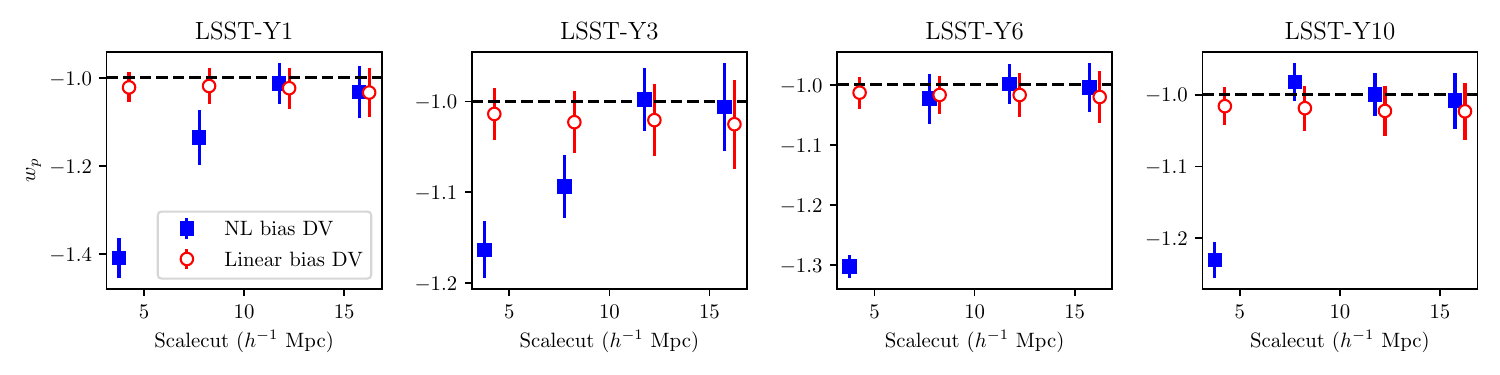}
    \caption{Impact of nonlinear galaxy bias on $w_p$. When analyzing linear bias data vector with a linear galaxy bias model (shown in red circles), there are no biases in the inference. Note however that due to projection effects, the resulting 1D posteriors are not centred on the fiducial value. However, when analyzing the data vector containing nonlinear galaxy bias with a linear bias model (shown in blue squares),  constraints are significantly biased at scale cuts $R_{\text{cut}}^{\text{gal}} \lesssim 8~h^{-1}$ Mpc.}
    \label{fig:galaxybias_impact_study}
\end{figure*}

\section{Including small scales in 3$\times$2  analysis}\label{sec:small_scales}

Our ability to include small scales in 3$\times$2 data vectors is limited by uncertainties in modeling baryonic physics and nonlinear galaxy bias. 
While discarding small scale data points ensures unbiased results, it also discards meaningful cosmological information. 


We explore scale cuts different from those we adopted from the DESC-SRD (see Section \ref{ssec:scale_cuts}) and quantify how they affect cosmological parameter constraints. We first conduct so-called impact studies to explore potential biases that occur when pushing to smaller scales (Section \ref{ssec:impact_studies}). After determining a range of appropriate scale cuts for our LSST scenarios, we explore the potential information gain/loss from choosing more aggressive/conservative scale cuts in Section \ref{ssec:small_scale_info}.

\subsection{Impact studies}\label{ssec:impact_studies}


In this section, we analyze synthetic data vectors that contain the effects of known systematics. Studying the robustness of our standard analysis pipeline with respect to these systematic effects allows us to determine acceptable scale cuts. We separately consider baryonic effects and nonlinear galaxy bias in Figure \ref{fig:baryon_impact_study} and Figure \ref{fig:galaxybias_impact_study}, respectively, using shifts in the 1D $w_p$ dimension as a metric. We consider a scale cut acceptable if the difference between the uncontaminated and contaminated data vector is $<0.3\sigma$ in $w_p$. Note, that this citeria is not equivalent to the parameter bias on $w_p$ itself being below this threshold due to the fact that projection effects cause 1D-biases that affect even the uncontaminated data vector (see black squares in Figure \ref{fig:baryon_impact_study} and red circles in Figure \ref{fig:galaxybias_impact_study}). 

In Figure \ref{fig:baryon_impact_study} we conduct our impact study when contaminating our data vectors with baryonic physics scenarios. As a first example, we look at the impact of baryons on dark energy constraints when analyzing the results with the optimistic scale cut of 1 arcmin in both $\xi_\pm$ (note that the fiducial scale cut was $\theta_{\text{min}} = 2.756 / 8.696$ arcmin for $\xi_{+/-}$). The results of this analysis are presented in Figure \ref{fig:baryon_impact_study}, where we plot the marginalized 1D posteriors of the dark energy parameter $w_p$ of the analysis with data vectors contaminated with baryonic scenarios of different hydrodynamical simulations. We analyze the data vectors with a varying number of baryon PCs and find that for most scenarios, marginalizing over 2 or 3 PCA modes is sufficient to capture the small-scale baryonic effects. We conclude that an aggressive scale cut of 1 arcmin in both parts of the cosmic shear data vector $\xi_{\pm}$ is acceptable for our simulated analyses.  

For our second example of impact study, we study the impact of nonlinear galaxy bias on 3$\times$2  analysis. In these simulated analyses, we contaminated our data vector with a more realistic nonlinear bias model using perturbation theory as described in \cite{Pandey2020}. In our simulated data vector, the second order bias parameter, $b_2$, is calculated in terms of the linear bias parameter, $b_1$, using the fitting formula of \cite{Lazeyras2016}. The tidal and the third-order bias term is set to their co-evolution values \citep{Saito2014, Pandey2020}. We analyze these data vectors for $4$ different scale cuts ranging from 4$h^{-1}$ Mpc to 16$h^{-1}$ Mpc. 

The results, shown in Figure \ref{fig:galaxybias_impact_study}, demonstrate that analyzing the nonlinear galaxy bias data vector with a linear bias model leads to substantial biases in the 1D posterior of $w_p$ for scale cuts $\lesssim 8~h^{-1}$ Mpc for Y1 and Y3 LSST scenarios. Based on this analysis we adopt 16 and 12 $h^{-1}$ Mpc as more agressive scale cuts that we explore further below.

Altogether these two figures show the results from 176 MCMC chains, a computational load that was massively sped up using one emulator. We note that biases that are more than 2$\sigma$ from the fiducial value are not credible given the accuracy of our emulator. For the purpose of determining scale cuts this does not matter, since 2$\sigma$ deviations are intolerable anyhow.


We note that in both cases, baryons and galaxy bias, the 1D parameter bias in other cosmological dimensions may exceed those of $w_p$. Which biases in which dimensions are tolerable for specific parameterizations and scale cuts needs to be decided by the DESC collaboration well before touching data.

\subsection{Information gain when including small scales}\label{ssec:small_scale_info}

\begin{figure*}
    \centering
    \includegraphics[width=\linewidth]{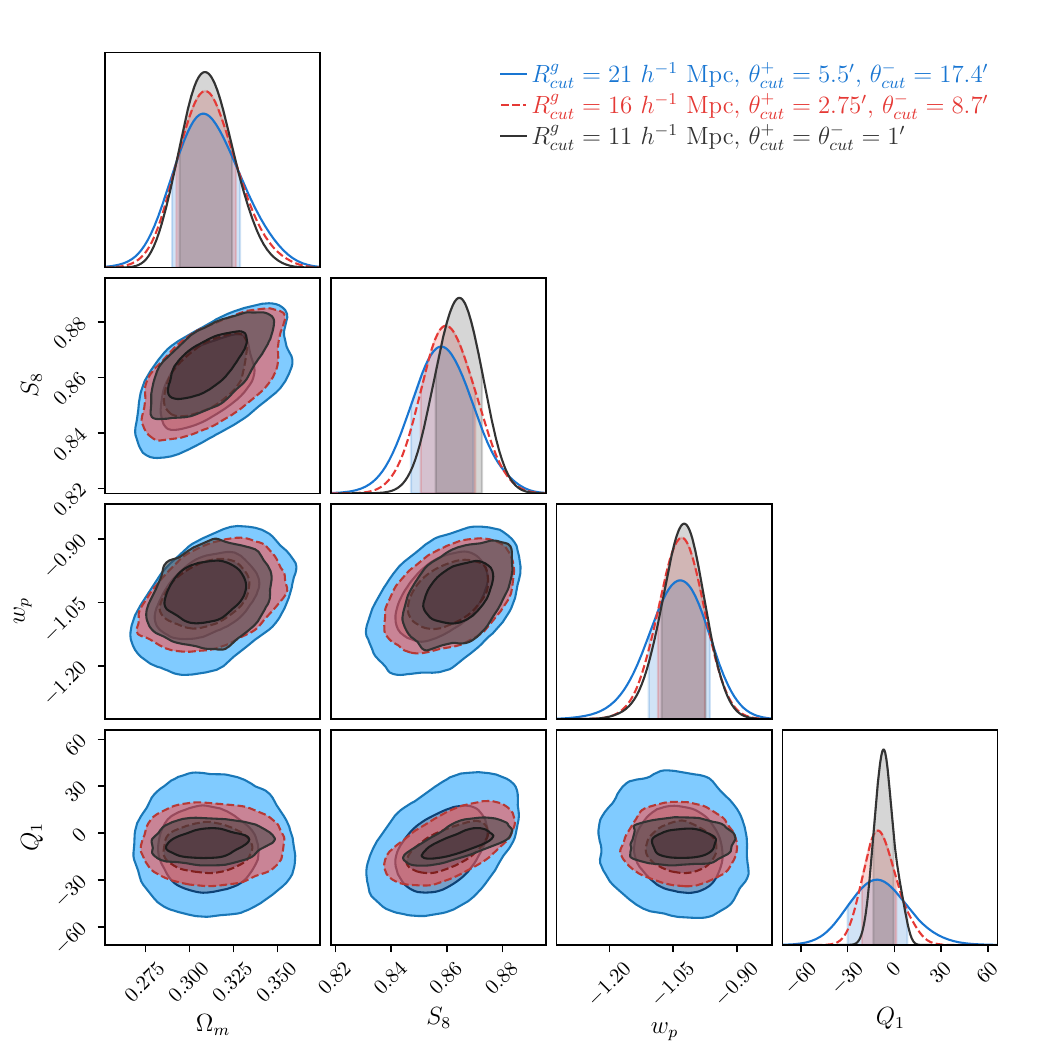}
    
    \caption{
    Constraints on cosmological and baryonic physics parameters obtained with LSST-Y1 3$\times$2 pt analysis using a conservative scale cut of $R^{\text{gal}}_{\text{cut}}=21~h^{-1}$ Mpc, $\theta^{+}_{\text{cut}}=5.5$ arcmin(blue), the baseline scale cut of $R^{\text{gal}}_{\text{cut}}=16~h^{-1}$ Mpc, $\theta^{+}_{\text{cut}}=2.7$ arcmin (red-dashed), and a more aggressive scale cut of $R^{\text{gal}}_{\text{cut}}=12~h^{-1}$ Mpc, $\theta^{+}_{\text{cut}}=1$ arcmin (gray). 
    Pushing to smaller scales improves the cosmological scales with the gain primarily coming from the small scales in GGL and galaxy clustering. Pushing to smaller scales also improves the constraints on baryonic physics as we can see by comparing the constraints on $Q_1$.
    }
    \label{fig:contour_plots_scalecut}
\end{figure*}

After validating new scale cuts we now move to quantify the possible information gain for the LSST Y1 scenario. We run cosmological analyses for the LSST-Y1 survey scenario with $3$ different scale cuts while marginalizing over all systematic parameters. 

For galaxy-galaxy lensing and galaxy clustering we assume scale cuts of $21/16/12~h^{-1}$ Mpc ranging from scales as conservative as in the DESC-SRD to the more aggressive range that we validated in the section above. We pair these scale cuts for 2$\times$2 with 3 choices for cosmic shear, where $\xi_+$ and $\xi_-$ get different scale cuts. From least to most aggressive our scale cuts read: $\theta^{+/-}_{\text{min}} = 5.5/17.4$ arcmin, $\theta^{+/-}_{\text{min}} = 2.75/8.7$ arcmin, $\theta^{+/-}_{\text{min}} = 1/1$ arcmin. We note that the middle scenario corresponds to the DESC-SRD choice.  

The results are shown in Figure \ref{fig:contour_plots_scalecut}, where blue corresponds to the most conservative, red to the intermediate, and black to the most aggressive scale cut. We see a marked improvement in the cosmological constraining power by pushing our analyses to smaller scales, which is something we strongly recommend DESC to consider. The dark energy figure-of-merit improves from $18.9$ to $26.4$ when going from the most conservative to the intermediate scale cut and to 31.5 for the most aggressive version. This is a significant gain and approaching the LSST Y3 FOM  of 38. 

We further note that the gain in cosmological information can be further enhanced by improved priors on the galaxy bias parameters. Our analysis indicates that the cosmological information gain primarily come from GGL and galaxy clustering parts of the data vector. 

Going to smaller scales in cosmic shear, significantly boosts the information on the baryon PC amplitude parameters (see the first one, $Q_1$, as an example in Fig. \ref{fig:contour_plots_scalecut}); the impact on cosmological parameters is limited.

\begin{figure}
    \centering
    \includegraphics[width=\linewidth]{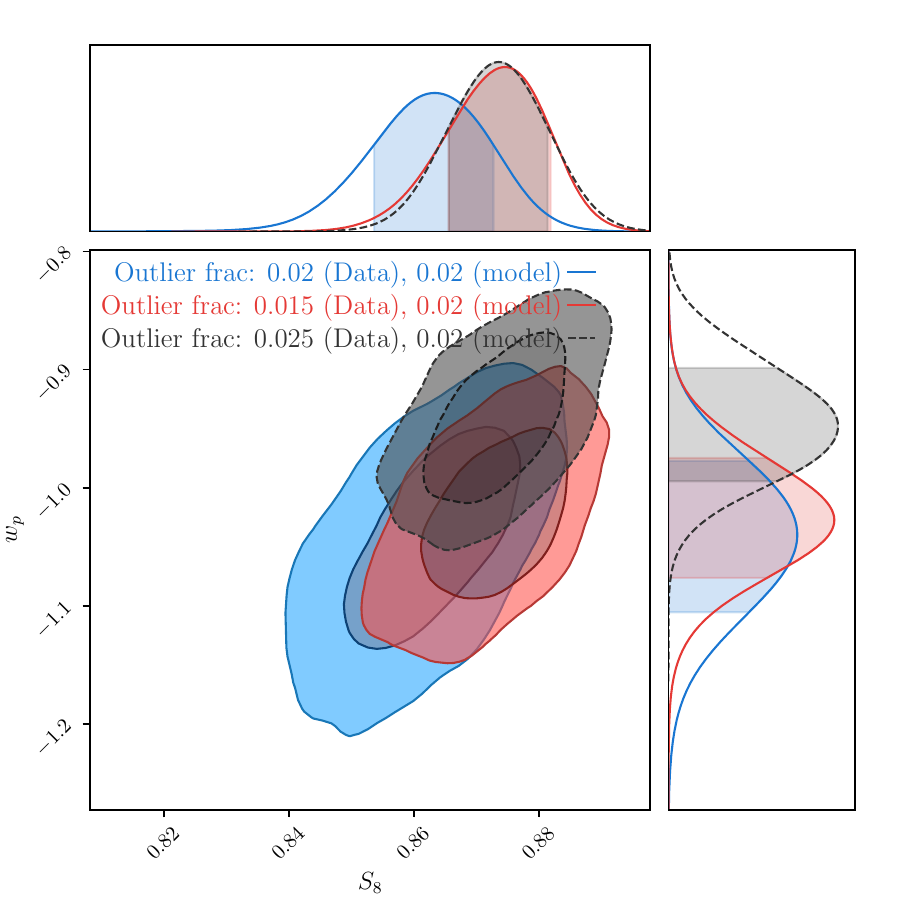}
    \caption{Impact of inaccurate determination of photo-$z$ outlier fractions on cosmological constraints with LSST-Y10 mock survey. For each of the analysis shown here, the model assumes the outlier fraction to be 0.02. But the simulated data vector is computed assuming an outlier fraction of 0.015 (blue contours), 0.02 (red contours) and 0.025 (grey dashed contours). As can be seen from the figure, inaccurate determination of the outlier fraction can lead to highly biased inference of the cosmological parameters.}
    \label{fig:outliers_contour_plot}
\end{figure}

\section{LSST synergies with spectroscopic datasets}\label{sec:spectro}
We explore synergies of LSST and spectroscopic datasets, e.g. obtained from the Dark Energy Spectrsocopic Instrument (DESI). These synergies manifest most prominently around two systematic classes of LSST, namely photo-$z$ uncertainties (Section \ref{ssec:photoz_outliers}) and intrinsic alignment ((Section \ref{ssec:lowz_specz}). 

We note that in the sections below we retrain the emulator if a parameter bias of more than 2$\sigma$ is observed in order to ensure the accuracy of all posteriors.

\subsection{Impact of catastrophic photo-$z$ outliers}\label{ssec:photoz_outliers}

In previous sections, we assume that uncertainty in the photometric redshift distributions for the tomographic bins can be accurately captured with just a constant shift. However, due to type/redshift degeneracies, some of the redshifts are mis-estimated by $\mathcal{O}(1)$ or more, commonly known as catastrophic redshift outliers. If these outliers are not properly addressed in the analysis, they can significantly skew the results of cosmological studies for Stage-IV surveys \citep{Sun2009, Bernstein2010, Hearin2010, Schaan2020, Fang2022}.

By conducting impact studies, we determine the level of accuracy required in controlling the outlier fraction to ensure that the cosmological analysis remains unbiased. We apply this analysis to our mock surveys for LSST-Y1 and Y10. The results of these analyses can be used to establish the spectroscopic requirements for a reliable cosmological analysis with LSST. 


To conduct our impact studies, we use the LSST mock galaxy sample created by \cite{Graham2018} that includes realistic photo-$z$ imperfections including catastrophic outliers. We fit a multivariate Gaussian to the joint probability distribution, $p_{\text{out}}(z_{\text{true}}, z_{\text{phot}})$, of the outlier sample and add it to the analytic redshift distribution (equation \eqref{eqn:smail}). We create a set of simulated data vectors with different outlier fractions and analyze them with an emulator trained with a fixed outlier fraction of $2\%$ (the outlier fraction in \cite{Graham2018}). 

With this analysis, we determine the accuracy required to estimate the outlier fraction to get robust cosmological constraints. This can be used to set requirements on the spectroscopic needs for LSST \citep{Sun2009, Bernstein2010}. It is worth noting that we do not consider cross-correlations between different redshift bins and the effect they may have on self-calibrating photo-$z$ outliers \citep{Schaan2020}.

\begin{figure}
    \centering
    \includegraphics[width=\linewidth]{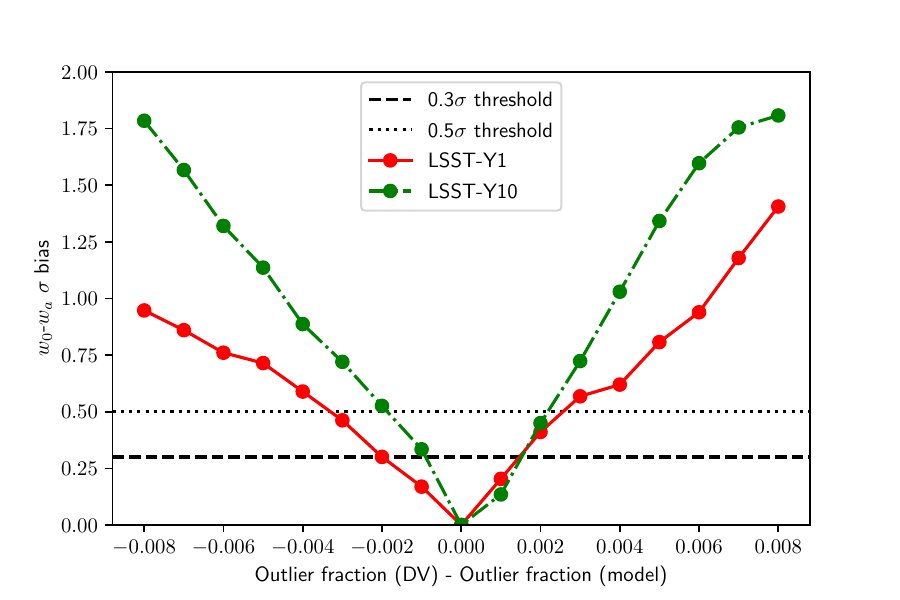}
    \caption{Bias in the $w_0$-$w_a$ parameters as a function of difference in the assumed outlier fraction for the LSST-Y1 (red circles) and LSST-Y10 (green circles) mock surveys. The dashed and dotted black lines corresponds to $0.3\sigma$ and $0.5\sigma$ bias in the $w_0$-$w_a$ plane. As we can see from the figure, in order to achieve cosmological inference with less than $0.5\sigma$ bias, we need to determine the outlier fraction to within $5\%$ ($10\%$) accuracy for LSST-Y10 (Y1).}
    \label{fig:outliers_tolerance}
\end{figure}

The results of our analysis are displayed in Figures \ref{fig:outliers_contour_plot} and \ref{fig:outliers_tolerance}. Figure \ref{fig:outliers_contour_plot} illustrates the impact on cosmological constraints if the fraction of outliers is mis-estimated in the analysis. The figure shows that a mis-estimation of the outlier fraction can result in significant biases in the $w_p$-$S_8$ plane. 

Figure \ref{fig:outliers_tolerance} shows the bias in the $w_0$-$w_a$ plane for varying levels of mis-estimation in the outlier fraction for LSST-Y1 and LSST-Y10 mock surveys. The figure demonstrates that in order to ensure that the cosmological constraints are not shifted by more than $0.3\sigma$, the outlier fraction must be determined to an accuracy of $0.001~(0.002)$ for LSST-Y10 (Y1), or a $5\%(10\%)$ determination of the outlier fraction. These are stringent requirements on the accuracy with which the outlier fraction must be known in order to ensure robust cosmological constraints using LSST 3$\times$2  analysis.  

\subsection{Mitigating IA uncertainties with low-$z$ spectroscopic surveys}\label{ssec:lowz_specz}

A high-density spectroscopic galaxy sample at $z < 1$ \citep[e.g,][]{Schlegel2022} can be used to directly measure intrinsic alignment through position-shape cross-correlations \citep[e.g.,][]{Singh2015, Samuroff2022}. \cite{Krause2016} suggest that removing or strongly mitigating intrinsic alignment uncertainties at low-$z$ will be sufficient for an LSST like analysis. Below we explore this idea in the context of our 3$\times$2 lens=source analysis.  

\begin{figure}
    \centering
    \includegraphics[width=\linewidth]{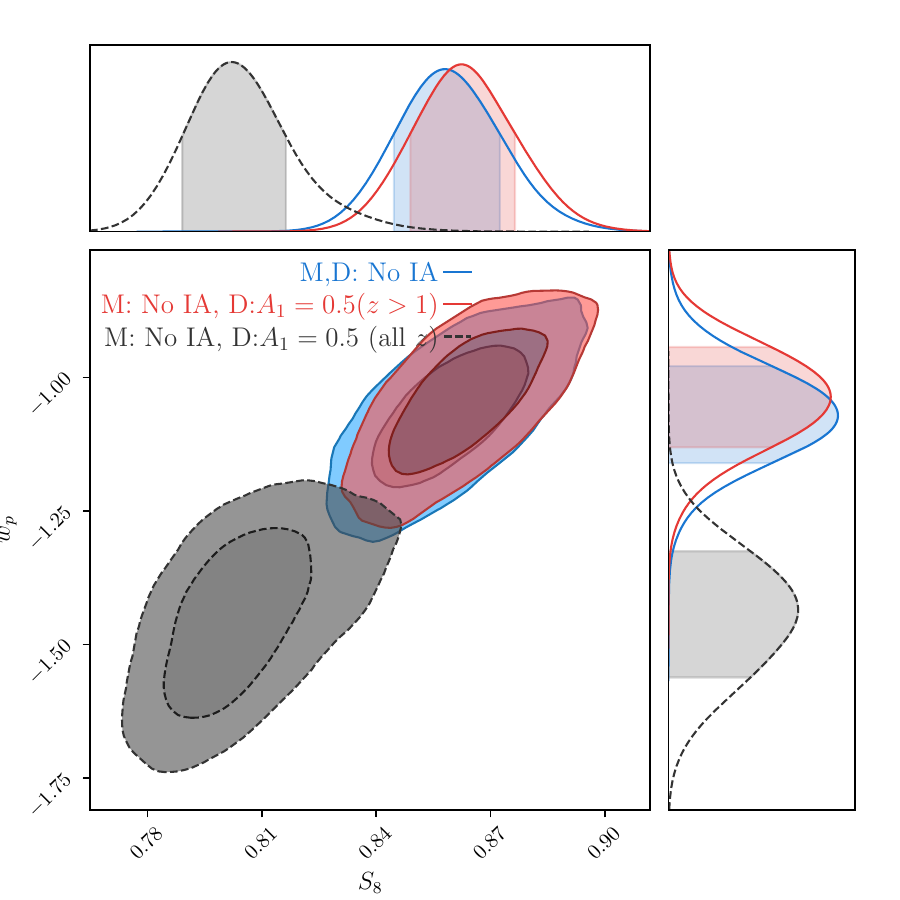}
    \caption{Impact of not modeling intrinsic alignment on cosmological constraints from LSST-Y1 3$\times$2 pt analysis. In all the three analyses, we ignore intrinsic alignment in the model. The blue contours show the analysis of a data vector that contains no intrinsic alignment contamination. The analysis of data vectors with IA contamination is shown with grey-dashed contours. As we can see, ignoring IA leads to large biases in the cosmological constraints. However, if we are able to perfectly model IA at $z < 1$ (mimicked by having IA contamination in the data vector only at $z > 1$), the resulting cosmological constraints are no longer biased (red contours). This result shows the value of a low-$z$ spectroscopic survey for mitigating IA in LSST analysis.}
    \label{fig:lowz_ia}
\end{figure}

We first create a synthetic data vector which contains our fiducial intrinsic alignment contribution and analyze it using a model with no IA. As we can see in Figure \ref{fig:lowz_ia}, ignoring IA in the analysis leads to a large bias in the cosmological constraints (grey contours). 

Under the assumption that a dense low-$z$ spectroscopic sample will be able to model the intrinsic alignment at low-$z$, we create a synthetic data vector that only has an IA contamination at $z > 1$.  When we analyze this data using a model with no intrinsic alignment, the resulting constraints in the $w_p$-$S_8$ plane (red contours) overlap almost perfectly with the constraints obtained if there was no IA in the universe (blue contours). 

We conclude that controlling IA at $z<1$ can largely solve the IA contamination for LSST, which is in line with the findings of \cite{Krause2016}.
Our analysis is optimistic in the sense that IA cannot be modelled perfectly even in the presence of a high-accuracy measurement with low-$z$ spectroscopy. The idea nevertheless is interesting and we leave a more detailed quantitative exploration to future work. 


While not shown here, we also investigate the impact these low-$z$ surveys can have on LSST 3$\times$2 pt analysis by better constraints on $\Delta_z$ at low redshifts. We find that better constraints on $\Delta_z$ can only marginally improve the cosmological constraints -- it leads to $\mathcal{O}(5\%)$ improvement in $w_0$-$w_a$ FoM.

\section{Summary}\label{sec:summary}

Understanding the response of the posterior probability to different analysis choices is a critical part of preparing for the data analysis of LSST (and other experiments). Running the required number of high precision MCMC chains is a significant computational bottleneck, which will only increase for more complex modeling analyses planned in the future. 

We introduced an iterative neural network based emulator design in \cite{Boruah2022} to solve this issue. The emulator allows us to run fast 3$\times$2 analyses while maintaining the realism of the full data analysis pipeline. 

In this paper, we used this emulator to study changes in parameter biases and error budgets for a `lens = source' 3$\times$2 analysis with LSST. Using the emulator allows us to run several hundreds of simulated likelihood analyses for LSST-Y1/Y3/Y6 and Y10 mock surveys while only using modest computational resources. While the specific results obtained in this paper will change for different analysis choices, the methodology described here is applicable for any analyses. 


For the analysis explored in this paper, we find that priors on galaxy bias parameters would provide the largest gain in cosmological information, whereas improved priors for baryonic or shear multiplicative uncertainties do not have a significant impact to the overall error budget. Between these two extremes, we find that constraints on photo-$z$ and IA uncertainties can improve the dark energy figure-of-merit noticeably, which directly points to synergies of future spectroscopic surveys and LSST.

We further explore different scale cuts than those suggested by the DESC-SRD in 3$\times$2 analyses by analyzing data vectors that are contaminated with known small scale systematics (galaxy bias and baryonic physics models). We determine the largest scalecut where the inferred cosmology is not significantly biased and subsequently run analyses with these more aggressive choices. We find significant gains in cosmological ($\sim 40\%$ improvement in the dark energy figure-of-merit) information with tolerable biases when adopting scale cuts that are smaller compared to those used in the DESC-SRD. 


     
Finally, we study synergies of LSST 3$\times$2 analyses with spectroscopic data sets in the context of a low redshift spectroscopic dataset as could be obtained from DESI. We explore two aspects in particular : {\it i)} the impact of catastrophic photo-$z$ outliers on LSST 3$\times$2 analysis and {\it ii)} intrinsic alignment mitigation strategies based on low-$z$ spectroscopic direct measurements of the effect. 


For the former, we find that the outlier fraction needs to be determined to within $5\%$ ($10\%$) accuracy to ensure unbiased inference of dark energy. Such analyses can be used to put requirements on the spectroscopic needs for photo-$z$ estimation. For our intrinsic alignment study, we find that mitigating IA at $z < 1$ leads to robust cosmological inference for LSST 3$\times$2  analysis -- presence of IA at high-$z$ do not noticeably bias cosmology parameters. 

To optimize future surveys like LSST but also Roman, SPHEREx, Euclid, DESI it is critical to map the consequences of different analysis choices. Our locally trained, iterative neural network emulator is a useful tool in this context, but the parameter space over which it can be trained sufficiently accurate is still small. We defer exploring more sophisticated neural network architectures that can span a larger parameter range to future studies. 



\begin{acknowledgments}
This paper has undergone internal review in the LSST Dark Energy Science Collaboration. 
The internal reviewers were Heather Prince, David Alonso and Georgios Valogiannis. We thank the internal reviewers for their useful comments. 
The computation presented here was performed on the High Performance Computing (HPC) resources supported by the University of Arizona TRIF, UITS, and Research, Innovation, and Impact (RII) and maintained by the UArizona Research Technologies department. This paper is supported by the Department of Energy HEP-AI program grant DE-SC0023892, and the Cosmic Frontier program grants DE-SC0020215 and DE-SC0020247.

Author contributions are as follows. SSB performed the main analyses and wrote the majority of the paper. TE supervised the research, designed the analyses and wrote the paper. VM is the main developer of COCOA and suggested analysis designs for testing the emulator. EF tested the implementation of baryons in COCOA. JM tested the COCOA pipeline and contributed to simulated analyses. EK suggested analyses for testing the impact of low redshift IA. XF is the main developer of CosmoCov and helped in the study of photo-z outliers. PR implemented and tested the non-linear galaxy bias implementation in COCOA.

The DESC acknowledges ongoing support from the Institut National de Physique Nucl\'eaire et de Physique des Particules in France; the Science \& Technology Facilities Council in the United Kingdom; and the Department of Energy, the National Science Foundation, and the LSST Corporation in the United States.  DESC uses resources of the IN2P3 Computing Center (CC-IN2P3--Lyon/Villeurbanne - France) funded by the Centre National de la Recherche Scientifique; the National Energy Research Scientific Computing Center, a DOE Office of Science User Facility supported by the Office of Science of the U.S.\ Department of Energy under Contract No.\ DE-AC02-05CH11231; STFC DiRAC HPC Facilities, funded by UK BEIS National E-infrastructure capital grants; and the UK particle physics grid, supported by the GridPP Collaboration.  This work was performed in part under DOE Contract DE-AC02-76SF00515.
\end{acknowledgments}

\bibliography{lsst_systematics}

\end{document}